\documentclass[colorlinks=true, linkcolor=blue, anchorcolor=blue, citecolor=blue, urlcolor=blue]{IEEEtran}
\usepackage{caption}
\usepackage{soul,framed} 
\usepackage[table]{xcolor}
\definecolor{navy}{RGB}{0, 0, 128}
\definecolor{steelblue}{RGB}{30, 144, 255}

\colorlet{shadecolor}{yellow}
\usepackage[pdftex]{graphicx}
\graphicspath{{../pdf/}{../jpeg/}}
\DeclareGraphicsExtensions{.pdf,.jpeg,.png}

\usepackage[cmex10]{amsmath}
\usepackage{amssymb}
\usepackage{stmaryrd}
\usepackage{physics}

\usepackage{array}
\usepackage{mdwmath}
\usepackage{mdwtab}
\usepackage{eqparbox}
\usepackage{url}

\usepackage{graphicx}
\usepackage{subcaption}
\usepackage{tabularx} 
\usepackage{graphicx}
\usepackage{amsmath}
\usepackage{booktabs}
\usepackage{color}
\usepackage{multirow}
\usepackage{colortbl}

\usepackage[numbers,sort&compress]{natbib} 
\usepackage{pifont}  
\usepackage{wasysym} 

\usepackage{mdwlist} 
 
\usepackage{algorithmic}

\usepackage[ruled,linesnumbered]{algorithm2e}
\usepackage{enumerate}
\usepackage{orcidlink}
\usepackage[export]{adjustbox} 

\def\BibTeX{{\rm B\kern-.05em{\sc i\kern-.025em b}\kern-.08em
    T\kern-.1667em\lower.7ex\hbox{E}\kern-.125emX}}

\usepackage{xpatch}
\xpatchcmd{\thebibliography}{\list}{\small\list}{}{}

\usepackage{pifont}
\usepackage{makecell}

\begin{document}

\bstctlcite{IEEEexample:BSTcontrol}
    \title{Sybil-TraceGuard: Traceability-enhanced Sybil Guardian for Connected and Autonomous Vehicles Using  Dynamic Semi-supervised GNN}
    
  \author{Qian~Xu,~\orcidlink{0000-0002-3374-4108}
      Jiaxun~Zhang,~\orcidlink{0009-0002-3234-2569}
      Chengyue~Wang,~\orcidlink{0009-0009-7707-1505}
      Zhenning~Li(Member),~\orcidlink{0000-0002-0877-6829}
      
  \thanks{Received Aug. 2026; (Corresponding author: Zhenning Li (zhenningli@um.edu.mo)}
  
  \thanks{Qian Xu, Jiaxun Zhang, Chengyue Wang, and Zhenning Li are with the State Key Laboratory of Internet of Things for Smart City, University of Macau, Macau SAR, China.Jiaxun Zhang, Chengyue Wang, and Zhenning Li are also with the Department of Civil and Environmental Engineering, Faculty of Engineering, University of Macau, Macau SAR, China. Zhenning Li is also with the Department of Artificial Intelligence, Faculty of Information Science and Computing, University of Macau, Macau SAR, China.}
   \thanks{This work was supported by the Science and Technology Development Fund of Macau [0007/2025/RIC, 0122/2024/RIB2, 0215/2024/AGJ,0074/2025/AMJ, 001/2024/SKL, 0002/2025/EQP], the Research Services and Knowledge Transfer Office, University of Macau [SRG2023-00037-IOTSC, MYRG-GRG2024-00284-IOTSC], the Shenzhen-Hong Kong-Macau Science and Technology Program Category C [SGDX20230821095159012], the Science and Technology Planning Project of Guangdong [2025A0505010016], National Natural Science Foundation of China [52572354], the State Key Lab of Intelligent Transportation System [2024-B001], and the Jiangsu Provincial Science and Technology Program [BZ2024055].}}


\markboth{IEEE TRANSACTIONS ON XXX, VOL.~X, NO.~X, ~X}{xxx \MakeLowercase{\textit{et al.}}s}

\maketitle


\begin{abstract}
Connected and autonomous vehicles (CAVs) face severe Sybil attacks, where attackers exploit privacy-preserving pseudonym-switching mechanisms to anomaly alternate identities while forging Basic Safety Messages (BSMs). 
Although existing schemes can flag suspicious behaviors, these temporally fragmented Sybil identities render traditional single-point and sequence-based deep learning methods ineffective. Linking these fragmented identities back to the source attacker is essential for root-cause elimination, particularly under extreme label scarcity. Therefore, the Sybil-TraceGuard is proposed as a dynamic semi-supervised spatio-temporal GNN framework for Sybil Guardian, prioritizing “who is responsible” over “whether an attack is happening”. It comprises four tightly coupled modules: Incremental Stream Attack Detection (ISAD) for efficient Sybil attack pre-screening; the Dynamic Topology-aware Constructor (DTC) for constructing spatio-temporal dynamic graphs; the Spatial GAT-Encoder with Multi-head Attention (SGEM) to capture multi-identity logical conflicts in spatial interactions; and the Multi-scale Spatio-Temporal Audit (MSTA) to audit short-term and long-term temporal inconsistencies. These modules are optimized within a semi-supervised Mean-Teacher framework via feature-edge shuffling perturbations, regularizing the latent feature space using minimal labels. Experiments across four Sybil attack scenarios demonstrate that Sybil-TraceGuard effectively links fragmented pseudonyms to source attackers. It outperforms state-of-the-art baselines across unlabeled ratios of 0.70–0.95, maintaining high stability and sensitivity despite extreme class imbalance and varying hyperparameter settings.

\end{abstract}

\textit{\textbf{Index Terms--}}
\textbf{Connected and autonomous vehicles, graph neural networks, semi-supervised learning, Sybil attack, source attacker traceability}

\IEEEpeerreviewmaketitle

\section{Introduction}

\IEEEPARstart{C}{onnected} and Autonomous Vehicles (CAVs)  rely on Vehicle-to-Everything (V2X) communication to improve road safety \cite{matin2022impacts}, traffic flow \cite{pan2024impacts} and environmental sustainability \cite{hua2026envir}. 
However, the openness and decentralized nature of V2X and profitable attack incentives also expose CAVs to various cybersecurity threats, including Denial of Service (DoS), kinematic-data manipulation, and Sybil attacks \cite{Bou2023ML, Abdel2025V2X}. Although the implementation of Sybil attacks varies across domains, the underlying logic remains consistent: it distorts the integrity of the consensus, thereby facilitating malicious activities. Among them, Sybil attacks for CAVs are particularly challenging because a single physical attacker can create multiple virtual identities and inject conflicting or falsified kinematic information \cite{Hammi2022Sybil, benarous2025Pse}. By continuously switching pseudonyms, an attacker can fragment its malicious behavior across seemingly independent identities, creating “illusion traffic” that disrupts cooperative awareness while concealing the physical source of the attacker.

\begin{figure}
    \begin{center}
        \includegraphics[width=3in]{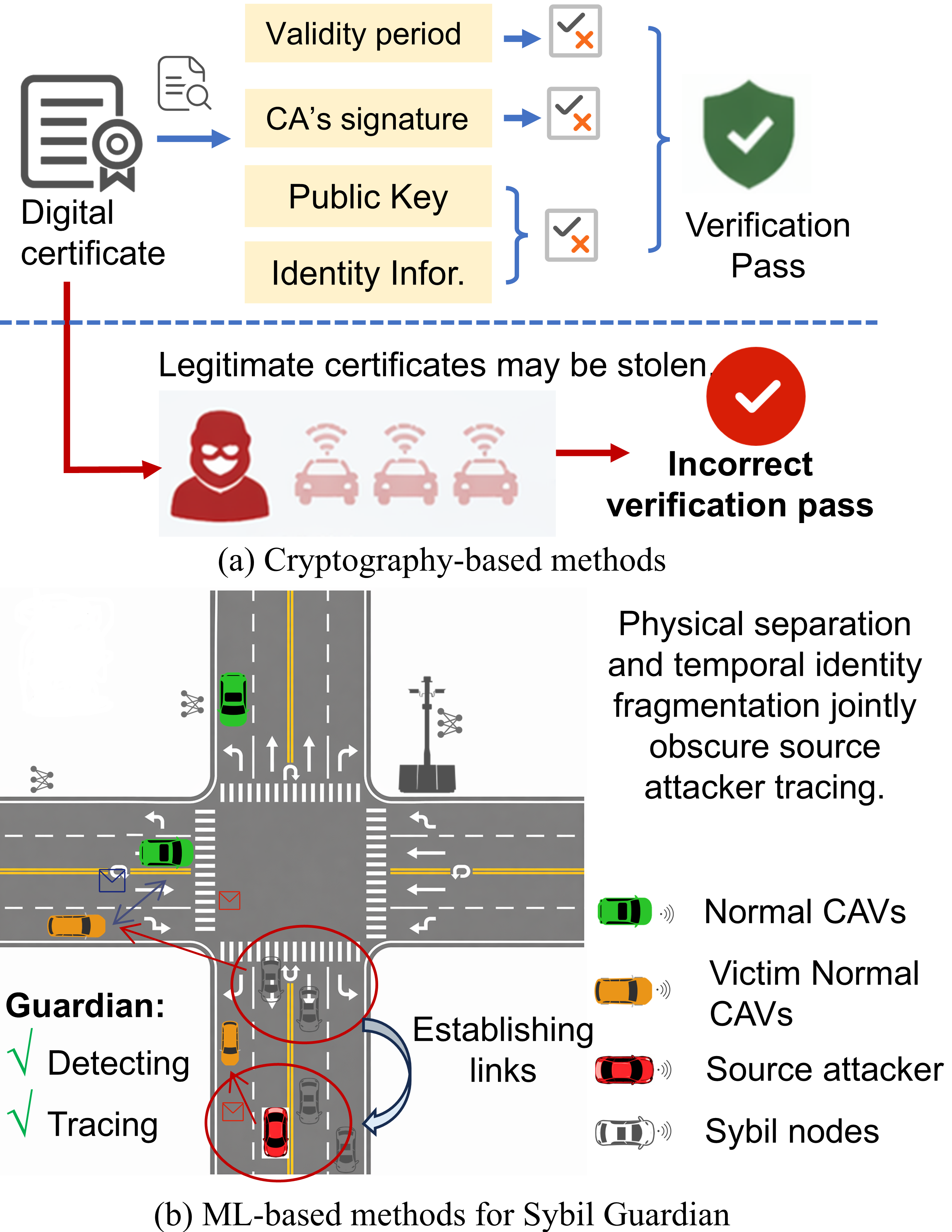}\\
        \caption{Comparison of Sybil attack detection approaches.}
        \label{fig:sybilmethod}
    \end{center}
\end{figure}

Existing Sybil detection and defenses are evolving from non-Machine Learning(ML) methods to ML methods, as well as  cross-layer defense and leveraging multiple data sources. Cryptographic authentication mechanisms may incur certificate-management overhead and remain vulnerable when legitimate credentials are stolen
\cite{baza2020Sybil}, as shown in Fig.~\ref{fig:sybilmethod}(a). Physical-layer indicators such as Received Signal Strength Indicator (RSSI) \cite{benadla2022RSSI} and Channel State Information (CSI) \cite{yao2018RSSI} are sensitive to environmental interference \cite{benadla2022RSSI,yao2018RSSI}. 
More recently, ML-based intrusion detection systems (IDSs) have leveraged V2X application-layer data besides network traffic data, particularly Basic Safety Messages (BSMs), which provide valuable spatio-temporal evidence connecting digital identities with physical motion \cite{SAEJ2735_2024}. 

While ML-based IDSs for CAVs have received increasing attention, existing studies have limitations in addressing the distinctive characteristics of Sybil attacks. Several novel frameworks fail to cover Sybil attacks \cite{Liu2026DTIDS, wang2026class}, while many approaches adopt a “one-size-fits-all" detection scheme \cite{xu2023secure, chen2024fast, aishwarya2026LLM}, failing to account for the unique stealthiness and spatio-temporal coupling inherent to Sybil attacks. Furthermore, sequence-based deep learning models, such as Convolutional Neural Networks (CNNs) or Long Short-Term Memory (LSTM) networks, are limited in capturing subtle yet critical spatio-temporal inconsistencies in Sybil attacks. Instead, Graph Neural Networks (GNNs) excel at modeling non-Euclidean topologies through message-passing mechanisms \cite{Bin2026GNNSurvey}, providing new thoughts for modeling Sybil attacks. However, improved detection alone remains insufficient, since once a suspicious pseudonym is blocked, the same physical attacker can discard it and continue the attack under another identity. Therefore, effective Sybil defense requires not only detecting malicious identities, but also tracing fragmented pseudonyms back to their physical source.

Achieving source attacker traceability from dynamic V2X streams presents three key challenges.

\textbf{(a) Limited traceability of source attackers.} 
Current ML-based detection paradigms focus on identifying attack occurrences \cite{Liu2026DTIDS, wang2026class,xu2023secure,chen2024fast,aishwarya2026LLM}, while the physical attacker remains decoupled from the virtual Sybil nodes it generates. Consequently, identity-level detection provides only temporary mitigation, necessitating a shift from attack occurrence detection toward persistent source attacker traceability.

\textbf{(b) Inadequate dynamic spatio-temporal representation via standard GNNs.} 
Tracing fragmented pseudonyms requires preserving behavioral consistency across time despite rapidly changing identities and communication topologies. However, conventional GNN formulations often process dynamic V2X streams as discrete graph snapshots, making it difficult to associate fragmented behaviors across evolving nodes and edges\cite{Feng2026DyGNN}. A traceability model should therefore jointly capture dynamic topology, spatial interactions, and temporal behavioral consistency.

\textbf{(c) Heavy reliance on high-quality labeled samples} 
Most existing attack detection methods depend on supervised learning algorithms \cite{Liu2026DTIDS}, \cite{wang2026class},\cite{xu2023secure}, \cite{aishwarya2026LLM}. 
In practice, obtaining fine-grained labels for evolving Sybil identities and their physical sources is costly, while newly emerging attack variants may further reduce the effectiveness of fully supervised models \cite{song2022graph},\cite{mvula2024SSL}. This motivates a novel semi-supervised GNN learning that can exploit large-scale unlabeled V2X streams while maintaining reliable traceability.

To address these challenges, we propose Sybil-TraceGuard, \textbf{trace}ability-enhanced \textbf{Sybil} \textbf{Guard}ian for CAVs using semi-supervised dynamic GNN. Following privacy-preserving V2X principles, the objective is to trace the responsible On-Board Unit (OBU), rather than reveal the driver's real-world identity.  The main contributions are summarized as follows.

\begin{itemize}
    \item \textbf{Innovation in detection paradigm.} 
     We formulate Sybil defense as a source attacker traceability problem, shifting the objective from identifying “whether an attack is happening” to determining “who is responsible”. Application-layer BSM evidence is exploited to associate fragmented pseudonyms through progressive behavioral consistency auditing.
     
    \item \textbf{Dynamic spatio-temporal GNN representation.} We propose a more discriminative feature space and a collaborative framework comprising four specialized modules.
    These modules are designed as follows. (i) an efficient online pre-screening module to filter out massive traffic before graph construction. (ii) a module to build and update the graph topology in dynamic environments, (iii) a module to analyze the spatial patterns, (iv) a module to analyze multi-scale temporal patterns.
    
    \item \textbf{Label-efficient semi-supervised traceability.} 
    We introduce a Mean-Teacher-based learning strategy with feature- and topology-level perturbations and dynamically weighted consistency regularization, allowing extensive unlabeled V2X data to contribute to representation learning. Experiments across four Sybil attack scenarios demonstrate robust traceability under severe label scarcity and diverse attack behaviors.
    
\end{itemize}

Our work is organized as follows: Section II reviews related work on Sybil attack detection and semi-supervised learning. Section III defines the system model and problem formulation. Section IV details the Sybil-TraceGuard methodology. Section V provides the experimental results, sensitivity analysis and ablation experiment. Finally, Section VI concludes the paper and discusses future directions.

%
\section{Related Works}
\subsection{Sybil Attack Detection and Traceability for CAVs}
Sybil attacks have observed in online social networks \cite{Zhang2023OSN},\cite{Al2017OSN}, blockchain systems \cite{Sybil2025Blockchain}, peer-to-peer networks \cite{patel2025survey}, federated learning \cite{Ya2024fedSybil}, etc. Existing taxonomies span position verification, resource testing, and data-driven reputation systems \cite{Hammi2022Sybil}, which were further expanded to encompass RSSI, cryptography, and trust and ML-based methods \cite{tang2025deep}. This section focuses on ML-based paradigms while briefly reviewing non-ML methods.

\textbf{Non-ML methods.}
Early Sybil defenses mainly relied on identity verification and physical
consistency. Baza et al. \cite{baza2020Sybil} integrated Proof-of-Location (PoL) and Proof-of-Work (PoW) and maximum-clique graph analysis to identify Sybil nodes.
Benadla et al. \cite{benadla2022RSSI} combined RSSI and blockchain-based PoL to trace conflicting trajectories. More recent studies have moved beyond identity-level detection toward source tracing through behavioral trajectory matching \cite{sultana2024coop} or beacon and neighborhood analysis \cite{zhu2024sybil}.

\textbf{Tabular ML-based Method.}
ML-based Sybil detection has evolved from feature-based classifiers to deep learning models. Representative approaches include Bayesian-optimized Random Forest using physical-layer features \cite{Abdel2025Sybil}, CNN/CNN-LSTM models for behavioral detection \cite{sultana2024Sybil}, and Gradient Boosting Decision Trees(GBDT)-based methods using BSM and traffic-flow information \cite{Chen2022Sybil}.
Most of these methods only focus on detecting malicious identities. A notable exception is \cite{chen2025sybil}, which adopts a two-stage GBDT binary classifier that first distinguishes Sybil nodes from real vehicles, and then further classifies real nodes into attackers and normal vehicles.

\textbf{Graph-based and GNN-based Methods.}
Graph-based Sybil attack detection was initially explored through probabilistic graph
inference and random-walk propagation in OSNs, such as SybilBelief \cite{Gong2014} and Sybil\_SAN \cite{Zhang2023SybilRW}. The inherently relational structure of V2X networks has subsequently
motivated research for CAVs.Luo et al. \cite{Luo2021Sybil} proposed a Credibility-Enhanced Temporal Graph Convolutional Network (TGCN) for Sybil attack detection, validated with a SUMO-generated dataset.
Tang et al. \cite{tang2025deep} proposed a supervised learning detection framework based on GCN and Gated Recurrent Unit (GRU). Nevertheless, few studies, apart from \cite{chen2025sybil}, validated the distinction between source attackers and Sybil nodes in independent experiments. Many studies applied classical ML methods and GNNs on limited Sybil attack variants, and most used supervised learning.

\subsection{Unsupervised and Semi-supervised ML for Anomaly Detection}
Semi-supervised learning (SSL) mitigates label scarcity by exploiting unlabeled samples together with limited supervision \cite{Yang2023SemiSurvey}. Early efforts at SSL-based anomaly detection mainly relied on statistical heuristics and clustering-based frameworks, such as self-training and multi-view co-training. Kristianto et al. ~\cite{kristianto2023misbehavior} combined federated semi-supervised learning
with pseudo-labeling and entropy minimization for misbehavior detection, including Sybil attacks. However, conventional SSL methods are primarily designed for batch learning and do not explicitly address continuously evolving V2X streams and graph structures.

For streaming anomaly detection, incremental clustering provides an efficient alternative to batch retraining. Micro-cluster-based frameworks, such as Density-based Clustering over an Evolving Data Stream (DenStream) \cite{cao2006denstream}, maintain time-decayed micro-clusters through fading and pruning mechanisms to adapt to evolving distributions. Algorithms like Density-Based Clustering in Data Streams (DBSTREAM) \cite{hahsler2016clustering} model the shared density between micro-clusters to capture arbitrary shapes. Such methods are suitable for high-throughput online pre-screening, but clustering alone provides limited capacity for fine-grained relational source traceability.

Recent SSL methods increasingly rely on consistency regularization. Mean-Teacher framework \cite{river_doc_overview_2026} uses Gaussian ramp-up weighting to enhance representation robustness across extensive unlabeled V2X streams. 
To address data imbalance, methods like FreeMatch \cite{river_doc_overview_2026} employ adaptive thresholding to refine label propagation. More recently, the integration of SSL with deep spatio-temporal architectures has allowed for the modeling of intricate structural dependencies. Duan et al. \cite{duan2022app} introduced a semi-supervised IDS using a dynamic line GNN (DLGNN) for
spatio-temporal intrusion detection.
Song et al. \cite{song2022graph} proposed a taxonomy of graph-based SSL, including transductive learning, inductive learning, and scalable Learning. 
Tian et al. \cite{tian2023sad} developed a semi-supervised anomaly detector combining temporal memory with
pseudo-label contrastive learning on dynamic graphs. 
Ekle et al. \cite{ekle2024anomaly} proposed a taxonomy of anomaly detection in dynamic graphs. Despite these advancements, existing graph-based SSL methods are not specifically designed to associate temporally fragmented Sybil identities with their physical sources under dynamic V2X topology.

\section{System Model}
\subsection{Attack Model}

\subsubsection{Attacker Capabilities and Assumptions} We assume an insider attacker as follows. 
\begin{itemize}
    \item \textbf{Identity legitimacy}: The attacker possesses valid pseudonym certificates and cryptographic keys, enabling them to bypass standard network-layer authentication.
    \item \textbf{Data manipulation}: The adversary exerts full control over the BSM, allowing for the arbitrary forging of position, velocity, and acceleration. 
    \item \textbf{Physical-layer constraints}: Despite masquerading as multiple logical identities, all Sybil messages originate from a single physical Radio Frequency (RF) chip. Consequently, the transmission process is subject to shared computational resources.
\end{itemize}

\subsubsection{Node Classification}
The nodes $\mathcal{V}$ are categorized into three types. 
\begin{itemize}
    \item \textbf{Normal CAVs ($V_{normal}$)}: Legitimate physical vehicles that strictly adhere to V2V protocols and broadcast accurate kinematic data.
    \item \textbf{Source Attackers ($V_{att}$)}: The physical malicious entities that orchestrate attacks by generating multiple forged identities.
    \item \textbf{Sybil Nodes ($V_{Sybil}$)}: Virtual identities fabricated by $V_{att}$. They exist only in the application layer and lack a physical counterpart on the road.
\end{itemize}

A normal CAV may become a victim when its situational awareness is affected by malicious messages. Therefore, victim vehicles represent a dynamic role rather than a separate node class, with $\mathcal{V}_{vic}^{t}\subseteq\mathcal{V}_{normal}$. $V_{att}$ and $V_{Sybil}$ are also called malicious vehicles $V_{mali}$.

\subsubsection{BSM Formulation}
A BSM is a spatiotemporal snapshot broadcast periodically between vehicles to facilitate cooperative awareness. The BSM state vector of node $i$ at time $t$, denoted as $\mathbf{x}_i^t$, is defined as Eq.~\eqref{eq:x}.
\begin{equation}
\label{eq:x}
\begin{aligned}
\mathbf{x}_i^t &= [ \mathbf{Pos}_i^t, \mathbf{Spd}_i^t, \mathbf{Acl}_i^t,\mathbf{Hed}_i^t, 
\\
&\sigma_i^t,\text{pseudo}_{i}^t, SendTime_i^t]^\top
\end{aligned}
\end{equation}
where $\mathbf{Pos}_i^t$, $\mathbf{Spd}_i^t$, $\mathbf{Acl}_i^t$, and $\mathbf{Hed}_i^t$ denote the reported position, velocity, acceleration, and heading vectors of node $i$ at time step $t$, respectively; $\sigma_i^t$ represents the measurement uncertainty; $\text{pseudo}_{i}^t$ is the active pseudonym of node $i$ at time step $t$; and $SendTime_i^t$ is the synchronized sender transmission timestamp.

Individual BSM observations can be forged or replayed, yet the attack behaviors inevitably manifest coupled anomalies at higher systemic levels, such as across fields, pseudonyms, relational contexts, and multiple time steps.

\subsubsection{Types of Sybil Attacks}
We consider four representative Sybil attack variants adopted from \cite{kamel2020MDS}. These attacks cover complementary adversarial strategies, including coordinated identity fabrication, high-rate randomized injection, legitimate-message imitation, and temporal replay, and therefore provide diverse behavioral patterns for evaluating Sybil defense. For clarity, each attack is characterized by its
“objective--strategy--anomaly” pattern.

\textbf{A1-Grid Sybil attack:} It aims to mislead neighboring vehicles by simulating traffic congestion, potentially causing planning confusion. A malicious vehicle generates a coordinated cluster of multiple virtual entities. By reducing the beacon interval, the attacker increases the overall message injection rate, successfully injecting multiple fake identities into neighbors’ tables within a single transmission cycle. In the self-grid mode, the cluster is centered on the attacker; in the remote-grid mode, the cluster is positioned at a detected neighbor’s location. Each virtual entity maintains a fixed relative formation within the cluster.

\textbf{A2-DoS Random Sybil attack:} It aims to exhaust the network and computational resources. A malicious vehicle disrupts the network by broadcasting high-frequency messages with randomized kinematic data under multiple identities. It significantly increases the transmission rate and requests a new pseudonym for each message.

\textbf{A3-DoS Disruptive Sybil attack:} 
It aims to bypass conventional identity checks while exhausting network and computational resources. A malicious vehicle captures legitimate messages from neighbors and rebroadcasts them at an accelerated rate using forged identities, with motion vectors that closely imitate normal traffic. 

\textbf{A4-Data Replay Sybil attack}: It aims to interfere with the temporal tracking of neighboring vehicles and introduce cumulative errors in trajectory prediction. A malicious vehicle captures authenticated messages from legitimate nodes and re-injects them into the network at a later time using forged identities. It consists of two stages: the attacker first replays realistic trajectories to maintain short-term realism, then switches to randomized trajectories once a predefined replay threshold is exceeded.

These variants are used as benchmark scenarios rather than attack-specific assumptions. The considered traceability problem targets behavioral inconsistencies across identities, message states, relational contexts, and temporal evolution, and is therefore not formulated around signatures of a particular attack type.

\begin{figure*}
    \begin{center}
        \includegraphics[width=6.5in]{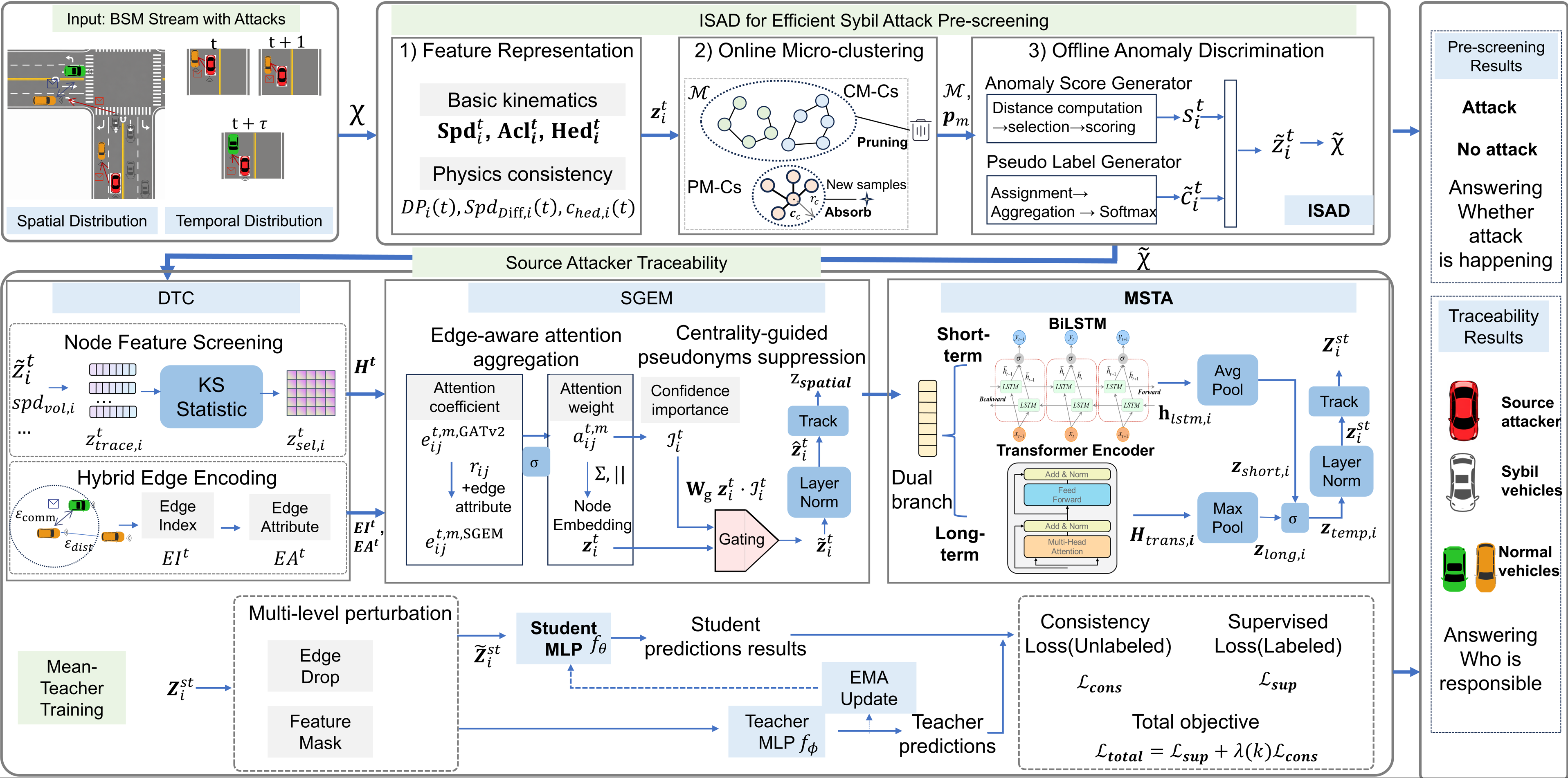}\\
        \caption{Overall framework of Sybil-TraceGuard. Stage I performs label-calibrated incremental Sybil attack pre-screening, while Stage II performs source attacker traceability through dynamic semi-supervised graph learning with DTC, SGEM, and MSTA.}
        \label{fig:methodframe}
    \end{center}
\end{figure*}

\subsection{Problem Formulation of Sybil-TraceGuard}

We formulate Sybil-TraceGuard as a two-stage online inference problem over dynamic V2X streams, aiming not only to detect suspicious Sybil behaviors but also to trace their persistent physical sources. At time $t$, the observed network is represented as $\mathcal{G}^{t}=(\mathcal{V}^{t},\mathcal{E}^{t},\mathbf{X}^{t})$, where $\mathcal{V}^{t}$ denotes logical identities, $\mathcal{E}^{t}$ their relational context, and $\mathbf{X}^{t}\in\mathbb{R}^{N_t\times D_{\mathrm{raw}}}$ the corresponding BSM observations. Over a temporal window of length $T$, the input is $\mathcal{G}_{\leq t} =\{\mathcal{G}^{\tau}\}_{\tau=t-T+1}^{t}$, whose identities and observations may evolve with pseudonym switching, mobility, and BSM manipulation.

Stage I performs label-calibrated incremental pre-screening: unsupervised stream clustering captures evolving behavioral patterns, while calibration labels from $\mathcal{D}_{\mathrm{cal}}\subseteq\mathcal{D}_{L}$ associate micro-clusters with binary detection semantics, yielding $\hat{a}_{i}^{t}\in\{0,1\}$. Stage II performs semi-supervised source attacker traceability over $\mathcal{G}_{\leq t}$:

\begin{equation}
\hat{y}_{i}^{t}
=
\mathcal{F}_{\theta}(\mathcal{G}_{\leq t}),
\qquad
\hat{y}_{i}^{t}\in\mathcal{C},
\end{equation}
where
$\mathcal{C}=\{\mathrm{Normal},\mathrm{Source\ Attacker},\mathrm{Sybil}\}$. The parameters $\theta$ are learned from sparse labeled data $\mathcal{D}_{L}$ and abundant unlabeled data $\mathcal{D}_{U}$, with $|\mathcal{D}_{L}|\ll|\mathcal{D}_{U}|$.

\section{Methodology}
\subsection{Overall Framework of Sybil-TraceGuard}
As illustrated in Fig.\ref{fig:methodframe}, Sybil-TraceGuard implements the two-stage formulation through an incremental pre-screening stage and a dynamic semi-supervised graph traceability stage.
In Stage I, the Incremental Stream Attack Detection (ISAD) module performs label-calibrated incremental pre-screening. Unsupervised micro-clustering with temporal fading captures evolving behavioral patterns, from which anomaly-aware representations and calibration-assisted binary pre-screening decisions are derived.
It filters normal traffic before graph construction, thereby reducing the computational burden of subsequent source attacker traceability.
In Stage II, the dynamic topology-aware construction (DTC) module constructs a hybrid dynamic graph. The Spatial GAT-Encoder with Multi-head Attention (SGEM) module and the Multi-scale Spatio-Temporal Audit (MSTA) module then capture spatial interactions and multi-scale temporal dynamics, respectively.
Under a Mean-Teacher-based SSL framework, an MLP classifier maps these representations into three classes: normal vehicles, source attackers, and Sybil nodes. This enables source attacker traceability under limited labeled data by exploiting abundant unlabeled observations.

\subsection{ISAD Module for Efficient Sybil Attack Pre-Screening}
ISAD adopts DenStream as its incremental clustering backbone and extends it with Sybil-oriented behavior representation and anomaly-aware representation augmentation for online pre-screening. Specifically, it comprises lightweight feature representation, online micro-clustering with temporal fading, and label-calibrated anomaly discrimination.

\subsubsection{Lightweight Feature Representation}
Given the raw BSM stream $\mathcal{X}$, we construct a lightweight feature representation $\mathbf{z}_i^t$ for node $i$ at time $t$ by concatenating basic kinematic features $\mathbf{z}_{\mathrm{kin},i}^t$ and short-term physics-consistency features $\mathbf{z}_{\mathrm{phy},i}^t$, where $\parallel$ denotes feature concatenation.
\begin{equation}
\label{eq:feature_space}
\mathbf{z}_i^t=\left[\mathbf{z}_{\mathrm{kin},i}^t\parallel\mathbf{z}_{\mathrm{phy},i}^t
\right],
\end{equation}

(i) Basic kinematic features: These features describe the instantaneous reported motion state of each node, which are extracted from the raw physical snapshots $\mathbf{x}_i^t$. Absolute position  $\mathbf{Pos}_i^t$ is excluded to avoid location-dependent bias and improve generalization across traffic environments.
\begin{equation}
\mathbf{z}_{\mathrm{kin}, i}^t = \left[ \mathbf{Spd}_i^t, \mathbf{Acl}_i^t, \mathbf{Hed}_i^t \right]^\top
\end{equation}

(ii) Physics consistency features: Since individual BSM states may be manipulated, instantaneous kinematics alone provide limited evidence of abnormal behavior. We therefore introduce three features to characterize cross-state discrepancies between consecutive BSMs.

\begin{equation}
\mathbf{z}_{\mathrm{phy},i}^t
=
\left[
DP_i(t),
Spd_{\mathrm{diff},i}^t,
c_{\mathrm{hed},i}^t
\right]^\top ,
\end{equation}

(a) The displacement prediction error ($DP$) measures the deviation between the reported position $\mathbf{Pos}_i^t$ and the position predicted from the previous state under a constant-velocity assumption. This lightweight prediction model is suitable for short-term consistency auditing over streaming BSMs, with $\Delta t$ denoting the inter-message sampling interval.

\begin{equation}
DP_i(t)=\left|\mathbf{Pos}_i^t-\left(\mathbf{Pos}_i^{t-1}+\mathbf{Spd}_i^{t-1}\Delta t\right)\right|_2
\end{equation}

(b) The speed difference ($Spd_{Diff}$) quantifies the short-term cross-field consistency between the position-derived speed $V_{pos}$ and the BSM-reported speed $V_{rep}$. Specifically, $V_{pos}$ is estimated from the positional displacement $\Delta d(t)$ between two consecutive reported positions, whereas $V_{rep}$ is obtained from the reported velocity vector from BSM:

\begin{subequations}
\label{eq:spd_consistency}
\begin{align}
Spd_{\mathrm{diff},i}^t&=\left|V_{\mathrm{pos},i}^t-V_{\mathrm{rep},i}^t\right|,
\\
V_{\mathrm{rep},i}^t&=\left\|\mathbf{Spd}_i^t\right\|_2,\qquad
V_{\mathrm{pos},i}^t=\frac{\Delta d_i^t}{\Delta t},
\\
\Delta d_i^t&=\left\|\mathbf{Pos}_i^t-\mathbf{Pos}_i^{t-1}\right\|_2 .
\end{align}
\end{subequations}

(c) The heading consistency ($c_{\mathrm{hed}}$) measures the consistency between the reported velocity
$\mathbf{Spd}_i^t$ and heading $\mathbf{Hed}_i^t$ using cosine similarity, thereby reducing the influence of vector magnitude:
\begin{equation}
\label{eq:c_hed}
c_{\mathrm{hed}, i}^t=\frac{(\mathbf{Spd}_i^t)^\top \mathbf{Hed}_i^t}{
\left\|\mathbf{Spd}_i^t\right\|_2\left\|\mathbf{Hed}_i^t\right\|_2+\epsilon
}.
\end{equation}
Here, $\epsilon$ is a small constant for numerical stability.

\subsubsection{Online Micro-Clustering Phase}
To adapt to evolving BSM streams without labeled samples, ISAD incrementally maintains time-decayed micro-clusters following the DenStream paradigm. The online state consists of potential micro-clusters (p-MCs), representing sufficiently supported behavioral patterns, and outlier micro-clusters (o-MCs), capturing sparse or emerging patterns. A fading factor $\lambda$ exponentially discounts historical observations, allowing recent behaviors to contribute more strongly to the evolving clustering structure.

For each incoming feature vector $\mathbf{z}_i^t$, ISAD first attempts to absorb it into the nearest compatible p-MC. If this fails, the sample is tested against the existing o-MCs. An o-MC is promoted to a p-MC once sufficient density support is accumulated; if no existing micro-cluster can absorb the sample, a new o-MC is initialized. Meanwhile, outdated micro-clusters gradually lose weight and may be pruned through temporal fading. Through continuous absorption, promotion, fading, and pruning, the online phase maintains the active micro-cluster state:
\begin{equation}
\label{eq:mt}
\mathcal{M}_t=\mathcal{M}_t^{p}\cup\mathcal{M}_t^{o}=\{(\mathbf{c}_m,r_m,w_m)\}_{m=1}^{M_t},
\end{equation}
where $\mathcal{M}_t^{p}$ and $\mathcal{M}_t^{o}$ denote the active p-MC and o-MC sets, respectively, and $\mathbf{c}_m$, $r_m$, and $w_m$ are the center, radius, and time-decayed weight of the $m$-th micro-cluster.

\subsubsection{Label-Calibrated Anomaly Discrimination}
The maintained micro-clusters capture the evolving density structure but do not directly provide anomaly semantics. ISAD therefore augments each observation with a cluster-relative anomaly score and a label-calibrated semantic prior. For $\mathbf{z}_i^t$, the anomaly score is derived from its nearest p-MC as:
\begin{subequations}
\label{eq:anomaly_score_calc}
\begin{align}
m_i^*&=\arg\min_{m\in\mathcal{M}_t^{p}}\frac{\|\mathbf{z}_i^t-\mathbf{c}_m\|_2^2}
{2r_m^2+\epsilon},
\\
s_i^t&=1-\exp\left(-\frac{\|\mathbf{z}_i^t-\mathbf{c}_{m_i^*}\|_2^2}{2r_{m_i^*}^2+\epsilon}
\right),
\end{align}
\end{subequations}
where $\epsilon$ ensures numerical stability, and larger $s_i^t\in[0,1)$ indicates stronger deviation from the established streaming pattern.

To associate the unsupervised clusters with binary detection semantics, a cluster-level prior is estimated exclusively from the labeled calibration set $\mathcal{D}_{\mathrm{cal}}$:
\begin{equation}
\label{eq:cluster_prior}
\mathbf{p}_m=\frac{\sum_{(i,\tau)\in\mathcal{D}_{\mathrm{cal}}}\mathbb{I}\!\left(\pi_\tau(\mathbf{z}_i^\tau)=m\right)\mathbf{y}_i^\tau
}{
\sum_{(i,\tau)\in\mathcal{D}_{\mathrm{cal}}}\mathbb{I}\!\left(\pi_\tau(\mathbf{z}_i^\tau)=m\right)
},
\end{equation}
where $\pi_\tau(\cdot)$ denotes micro-cluster assignment and $\mathbf{y}_i^\tau$ is the one-hot binary label. Hence, $\mathbf{p}_m$ is the calibration-derived normal--attack distribution of cluster $m$;
it remains fixed during inference, while clusters without calibration support use $\mathbf{p}_m=\mathbf{0}$.

During inference,
$\hat{\mathbf{c}}_i^t=\mathbf{p}_{\pi_t(\mathbf{z}_i^t)}$.
The binary pre-screening decision and anomaly-aware representation are defined as:
\begin{subequations}
\label{eq:augmented_representation}
\begin{align}
\hat{a}_i^t&=\mathbb{I}\left([\hat{\mathbf{c}}_i^t]_{\mathrm{attack}}\ge \eta\right),
\label{eq:prescreen_decision}
\\
\tilde{\mathbf{z}}_i^t&=\left[\mathbf{z}_i^t\parallel
s_i^t\parallel
\hat{\mathbf{c}}_i^t\right],
\label{eq:augmented_feature}
\\
\tilde{\mathcal{X}}_{\leq t}&=
\left\{\tilde{\mathbf{z}}_i^\tau\mid
i\in\mathcal{V}^{\tau},\;
\tau=t-T+1,\ldots,t\right\},
\label{eq:augmented_stream}
\end{align}
\end{subequations}
where $\eta\in[0,1]$ is the decision threshold. The augmented stream
combines behavioral features, anomaly evidence, and calibration-derived
semantics for subsequent graph-based traceability.

\subsection{Dynamic Semi-supervised GNN for Source Attacker Traceability}
\subsubsection{DTC Module for Dynamic Topology Construction}
Conventional graph construction methods based on basic node features and physical proximity may overlook behavioral anomalies related to source attacker traceability. DTC module therefore constructs a hybrid dynamic graph by incorporating traceability-oriented node feature screening with communication- and physical proximity-based edge encoding. To support efficient graph processing, the topology is represented in sparse Coordinate (COO) format:
\begin{equation}
\mathcal{G}^t
=f_{\mathrm{DTC}}(\tilde{\mathcal{X}}_{\leq t})=\big\{
\mathbf{H}^t,\mathbf{EI}^t,\mathbf{EA}^t\big\}.
\end{equation}
where $\mathbf{H}^t\in\mathbb{R}^{N_t\times d_{\mathrm{sel}}}$ is the selected node feature matrix, $\mathbf{EI}^t\in\mathbb{N}^{2\times|\mathcal{E}^t|}$ is the directed sparse edge index combining communication and proximity relationships, and $\mathbf{EA}^t\in\mathbb{R}^{|\mathcal{E}^t|\times1}$ contains the corresponding continuous edge attributes.

\textbf{Traceability-oriented node feature screening.}
Sybil attacks may still preserve short-term consistency while exhibiting longer-term temporal and communication anomalies. DTC therefore extends the anomaly-aware representation $\tilde{\mathbf{z}}_i^t$ with statistical volatility and traffic-context features into the traceability feature candidate vector $\mathbf{z}_{\mathrm{trace}, i}^t$:

\begin{equation}
\label{eq:trace_feature_space}
\begin{aligned}
\mathbf{z}_{\mathrm{trace},i}^t=\left[
\tilde{\mathbf{z}}_i^t\parallel\mathbf{z}_{\mathrm{vol},i}^t \parallel\mathbf{z}_{\mathrm{ctx},i}^t
\right],
\end{aligned}
\end{equation}

(i) \textit{Statistical volatility features.}
Speed volatility $spd_{\mathrm{vol},i}(t)$ and jerk volatility
$jerk_{\mathrm{vol},i}(t)$ characterize motion fluctuations over a
sliding window of length $W$. The former measures variations in the
reported speed magnitude, while the latter captures abrupt acceleration
changes:
\begin{subequations}
\label{eq:volatility_features}
\begin{align}
spd_{\mathrm{vol},i}(t)&=
\operatorname{std}\left(\left\{
V_{\mathrm{rep},i}(\tau)
\right\}_{\tau=t-W+1}^{t}\right),
\\
\mathbf{j}_i^\tau&=\frac{
\mathbf{Acl}_i^\tau-\mathbf{Acl}_i^{\tau-1}}{\Delta t},
\\
jerk_{\mathrm{vol},i}(t)&=\operatorname{std}\left(\left\{
\|\mathbf{j}_i^\tau\|_2
\right\}_{\tau=t-W+2}^{t}
\right).
\end{align}
\end{subequations}
Since each jerk value requires two consecutive acceleration observations,
the jerk sequence contains $W-1$ valid samples.

(ii) \textit{Traffic-context features.}
The six traffic-context features capture communication activity, identity
persistence, pseudonym multiplicity, and concurrent spatial behavior that
are not directly reflected by motion observations. Specifically,
$fre_{\mathrm{msg},i}(t)$ is the average number of BSMs associated with
identity $i$ over the latest $W$ time steps. For a message $m$ with
identifier $id_m$, $id\_age(m)$ is the elapsed time since the identifier
first appeared, $id\_cum\_msg(m)$ is its cumulative message count up to
$t_m$, and $pseudo\_per\_id(m)$ is the number of distinct pseudonyms
associated with it up to $t_m$. Moreover, $node\_density(m)$ counts the
distinct active pseudonyms observed at time $t_m$. The spatial-overlap feature $spatial\_overlap(m)$ further counts the distinct pseudonyms reporting the same position as message $m$ at time $t_m$:
\begin{equation}
\label{eq:spatial_overlap}
\begin{aligned}
spatial\_overlap(m)&=\left|
\left\{
\mathrm{Pseudo}_{m'}
\mid m'\in\mathcal{M},\;
t_{m'}=t_m,\right.\right.\\ &\qquad\left.\left.\mathbf{Pos}_{m'}=\mathbf{Pos}_m \right\}
\right|.
\end{aligned}
\end{equation}

To accommodate heterogeneous Sybil patterns, DTC applies KS-based screening to each candidate feature $f\in\mathbf{z}_{\mathrm{trace},i}^t$. The score $\mathrm{KS}_f$ is defined as the maximum pairwise CDF discrepancy among Sybil nodes $F_{\mathrm{sybil}}(x)$, source attackers $F_{\mathrm{source\_att}}(x)$, and normal vehicles $F_{\mathrm{norm}}(x)$. The KS scores are
computed exclusively from the labeled training data, and the selected feature subset is fixed during validation and testing. Features exceeding the screening threshold are retained and stacked to form the node feature matrix:

\begin{subequations}
\label{eq:ks_feature_screening}
\begin{align}
\mathrm{KS}_f &=\max\left\{\begin{aligned}&\sup_x\left| F_{\mathrm{sybil}}^{f}(x)-F_{\mathrm{norm}}^{f}(x)\right|,
\\&\sup_x
\left|
F_{\mathrm{source\_att}}^{f}(x)-F_{\mathrm{norm}}^{f}(x)\right|,
\\
&\sup_x\left| F_{\mathrm{source\_att}}^{f}(x)-F_{\mathrm{sybil}}^{f}(x)\right|
\end{aligned}
\right\},
\label{eq:ks_statistic}
\\
\mathbf{z}_{\mathrm{sel},i}^t&=\left\{f\in\mathbf{z}_{\mathrm{trace},i}^t\mid
\mathrm{KS}_f>\theta_{\mathrm{KS}}\right\},
\label{eq:ks_selection}
\\
\mathbf{H}^t&=\left[\mathbf{z}_{\mathrm{sel},1}^t,\mathbf{z}_{\mathrm{sel},2}^t,
\dots,
\mathbf{z}_{\mathrm{sel},N_t}^t\right]^\top
\in
\mathbb{R}^{N_t\times d_{\mathrm{sel}}}.
\label{eq:ht}
\end{align}
\end{subequations}
where $\theta_{\mathrm{KS}}$ is the screening threshold.  $\mathbf{z}_{\mathrm{sel},i}^t$ is the ordered feature vector
retained for node $i$ at time $t$, and $d_{\mathrm{sel}}$ is the number
of selected features. 

\textbf{Hybrid edge encoding.}
\textit{(i) Edge Index.}
Communication edges ($\mathcal{E}_{\mathrm{comm}}$) encode explicit BSM receiver relationships, while distance-based edges ($\mathcal{E}_{\mathrm{dist}}$) connect nearby nodes to mitigate graph fragmentation caused by intermittent V2X communication:
\begin{subequations}
\label{eq:edge_index_construction}
\begin{align}
(i,j)\in\mathcal{E}_{\mathrm{comm}}^t&\iff\mathrm{Pseudo}_j\in\mathcal{R}_i^t,\label{eq:comm_edge}
\\
(i,j)\in\mathcal{E}_{\mathrm{dist}}^t&\iff
\left\|\mathbf{p}_i^t-\mathbf{p}_j^t\right\|_2<R_{\mathrm{thr}},
\label{eq:dist_edge}
\\
\mathcal{E}^t&=\mathcal{E}_{\mathrm{comm}}^t\cup\mathcal{E}_{\mathrm{dist}}^t.
\label{eq:final_edge_set}
\end{align}
\end{subequations}
where $\mathrm{Pseudo}_j$ denotes the pseudonym of node $j$,$\mathcal{R}_i^t$ is the receiver set of node $i$ at time $t$,$\mathbf{p}_i^t$ is its spatial position, and $R_{\mathrm{thr}}$ is the proximity threshold. The ordered pairs in $\mathcal{E}^t$ form the columns of $\mathbf{EI}^t$.

\textit{(ii) Edge Attribute:}  
Each edge $(i,j) \in \mathbf{EI}^t$ is assigned a continuous attribute to quantify interaction strength $r_{ij}^t$. $\mathbf{EA}^t$ encapsulates these values following the sequence of sparse node pairs.
\begin{equation}
\begin{aligned}
\mathbf{EA}^t&=\left\{r_{ij}^{t}\mid(i,j)\in\mathbf{EI}^{t}\right\},\qquad
\\
r_{ij}^{t}&=\frac{1}{\left\|\mathbf p_i^{t}-\mathbf p_j^{t}\right\|_2+1}
\end{aligned}
\end{equation}
where $\mathbf p_i^{t}$ and $\mathbf p_j^{t}$ denote the spatial positions of nodes $i$ and $j$ at time step $t$, respectively.

\subsubsection{SGEM Module for Spatial Logic Auditing}
Given $\mathcal{G}^t$, SGEM combines edge-aware attention aggregation and centrality-guided pseudonym suppression to obtain the spatial representation $\mathbf{Z}_{\mathrm{spatial}}$.

\textbf{Edge-aware attention aggregation.} 
Although $\mathrm{GATv2}$ improves attention expressiveness over conventional $\mathrm{GAT}$ against the static attention dilemma, it exclusively considers node features. 
However, Sybil attacks may induce coordinated anomalies in both node and spatial interactions. 
SGEM module therefore incorporates the continuous edge attributes $\mathbf{EA}^{t}$ into the dynamic attention computation. For a node pair $(i,j)\in\mathbf{EI}^{t}$ and the $m$-th attention head,

\begin{subequations}\label{eq:spatial_attention}
\begin{align}
e_{ij}^{t,m,\mathrm{GATv2}}&=\mathbf a_m^{\top}\mathrm{LeakyReLU}\left(\mathbf W_m\left[
\mathbf h_i^{t}\parallel\mathbf h_j^{t}\right]\right)
\label{eq:gatv2_base}
\\
e_{ij}^{t,m,\mathrm{SGEM}}&=\mathbf a_m^{\top}\mathrm{LeakyReLU}\left(
\mathbf W_m\left[\mathbf h_i^{t}\parallel\mathbf h_j^{t}\parallel
r_{ij}^{t}\right]\right)
\label{eq:sgem_edge_aware}
\end{align}
\end{subequations}

where $\mathbf h_i^{t},\mathbf h_j^{t}\in\mathbf H^{t}$ are the input node features, $r_{ij}^{t}\in\mathbf{EA}^{t}$ is the edge attribute, and $\mathbf a_m$ and $\mathbf W_m$ denote the attention vector and projection matrix of the $m$-th head, respectively.

The edge-aware coefficients are normalized over $\mathcal{N}(i)$ and aggregated across $H$ attention heads as follows.
\begin{subequations}\label{eq:spatial_aggregation}
\begin{align}
\alpha_{ij}^{t,m}&=\mathrm{softmax}_{j\in\mathcal{N}(i)}\left(e_{ij}^{t,m,\mathrm{SGEM}}\right),
\label{eq:spatial_softmax}
\\
\mathbf z_i^{t}&=\mathop{\parallel}_{m=1}^{H}\left(\sum_{j\in\mathcal N(i)}\alpha_{ij}^{t,m}
\mathbf W_m \mathbf h_j^{t}\right).
\label{eq:spatial_embedding}
\end{align}
\end{subequations}

where $\mathcal{N}(i)$ denotes the neighborhood of node $i$ and $H$ is the number of attention heads.

\textbf{Centrality-guided pseudonym suppression.}
To suppress weakly connected yet highly suspicious pseudonyms, SGEM measures the
incoming attention centrality of node $i$ as:

\begin{equation}
\mathcal{I}_{i}^{t} = \frac{1}{H} \sum_{j \in \mathcal{N}(i)} \sum_{m=1}^{H} \alpha_{ji}^{t,m}
\end{equation}

The centrality score gates the intermediate embedding $\mathbf{z}_i^t$, followed by residual projection and layer normalization, where $\odot$ denotes element-wise multiplication, and $\mathbf{W}_g$ is the trainable gating matrix. Finally, The resulting embeddings $\mathbf{Z}_{\mathrm{spatial}}$ over the tracking window $\mathcal{W}$ form.

\begin{subequations}\label{eq:centrality_gate}
\begin{align}
\tilde{\mathbf{z}}_{i}^{t}&= \mathbf{z}_{i}^{t} \odot \mathrm{Sigmoid} \left( \mathbf{W}_g \mathbf{z}_{i}^{t} \cdot \mathcal{I}^{t}_{i} \right)
\label{eq:gate_apply}
\\
\hat{\mathbf{z}}_{i}^{t}&=\mathrm{LayerNorm} \left( \mathrm{Proj}(\tilde{\mathbf{z}}_{i}^{t}) + \mathbf{h}_{i}^{t} \right)
\label{eq:residual_ln}
\\
\mathbf{Z}_{\mathrm{spatial}} &= \Big[ \hat{\mathbf{z}}_i^t \Big]_{i=1, \, t=1}^{N, \quad |W|} \in \mathbb{R}^{N \times |W| \times d}
\end{align}
\end{subequations}

where $N$ is the number of tracked nodes and $d$ is the hidden feature dimension.

\subsubsection{MSTA Module for Multi-Scale Temporal Consistency Auditing}
SGEM captures neighborhood-level spatial logic but does not explicitly model dependencies across consecutive frames. MSTA therefore employs parallel BiLSTM and Transformer branches to capture short- and long-range temporal dependencies, respectively. Formally, for vehicle $i$, its spatial representation sequence within the window $W$ is extracted as a matrix slice $\mathbf{Z}_{\mathrm{spatial}, i} = [ \hat{\mathbf{z}}_i^1, \dots, \hat{\mathbf{z}}_i^{|W|} ]^\top \in \mathbb{R}^{|W| \times d}$. This slice is linearly projected into a latent temporal space $\mathbf{H}_{\mathrm{base}, i} \in \mathbb{R}^{|W| \times d}$.
\begin{equation}
\mathbf{H}_{\mathrm{base}, i} = \mathbf{Z}_{\mathrm{spatial}, i} \mathbf{W}_{\mathrm{emb}} + \mathbf{b}_{\mathrm{emb}}
\end{equation}

where $\mathbf{W}_{\mathrm{emb}} \in \mathbb{R}^{d \times d}$ and $\mathbf{b}_{\mathrm{emb}} \in \mathbb{R}^d$ are trainable parameter matrices. 

\begin{equation}
\mathbf{Z}_{\mathrm{spatial},i}
=
\left[
\hat{\mathbf{z}}_i^\tau
\right]_{\tau\in\mathcal{W}}
\in
\mathbb{R}^{|\mathcal{W}|\times d},
\end{equation}

Subsequently, the temporal features are concurrently routed through the dual-path architecture to construct multi-scale structural constraints.
\begin{subequations}\label{eq:temporal_audit}
\begin{align}
\mathbf{h}_{\mathrm{lstm}, i}&=\mathrm{BiLSTM}(\mathbf{H}_{\mathrm{base}, i})
\label{eq:bilstm_path}
\\
\mathbf{z}_{\mathrm{short}, i} &= \mathrm{AvgPool}_t(\mathbf{h}_{\mathrm{lstm}, i})
\label{eq:short_term_descriptor}
\\
\mathbf{H}_{\mathrm{trans}, i}&=\mathrm{Transformer}(\mathbf{H}_{\mathrm{base}, i} + \mathbf{P})
\label{eq:transformer_path}
\\
\mathbf{z}_{\mathrm{long}, i} &= \mathrm{MaxPool}_t(\mathbf{H}_{\mathrm{trans}, i})
\label{eq:long_term_descriptor}
\end{align}
\end{subequations}

where $\mathbf{P}\in\mathbb{R}^{|\mathcal{W}|\times d}$ is a learnable positional encoding, and pooling is performed along the temporal dimension. The resulting $\mathbf{z}_{\mathrm{short},i},\mathbf{z}_{\mathrm{long},i} \in\mathbb{R}^{d}$ represent the short- and long-range temporal contexts, respectively.

The localized and global temporal profiles are fused via a non-linear bottleneck layer, followed by a residual shortcut connecting the coarse-grained temporal context to produce the spatio-temporal embedding vector $\mathbf{z}_i^{\mathrm{st}} \in \mathbb{R}^d$. By horizontally stacking the transposed vectors of all $N$ entities, the output feature matrix $\mathbf{Z}_{\mathrm{st}} \in \mathbb{R}^{N \times d}$ is derived.
\begin{subequations}\label{eq:temporal_fusion}
\begin{align}
\mathbf{z}_{\mathrm{temp}, i}&=Softmax \left( \mathbf{W}_f [ \mathbf{z}_{\mathrm{short}, i} \parallel \mathbf{z}_{\mathrm{long}, i} ] + \mathbf{b}_f \right)
\label{eq:fusion_layer}
\\
\mathbf{z}_i^{\mathrm{st}}&=\mathrm{LayerNorm} \left( \mathbf{z}_{\mathrm{temp}, i} + \mathrm{AvgPool}_t(\mathbf{H}_{\mathrm{base}, i}) \right)
\label{eq:final_fingerprint}
\\
\mathbf{Z}_{\mathrm{st}}&=\Big[ \mathbf{z}_i^{\mathrm{st}} \Big]_{i=1}^{N}
\label{eq:final_tensor_stack}
\end{align}
\end{subequations}
where $\parallel$ denotes feature concatenation, $\sigma(\cdot)$ is the activation function, and $\mathbf{W}_f \in \mathbb{R}^{2d \times d}$ and $\mathbf{b}_f \in \mathbb{R}^d$ are trainable weight parameters.

\subsubsection{Mean-Teacher Optimization Objective}
The $\mathrm{Mean}$-$\mathrm{Teacher}$ framework leverages sparse labels by enforcing prediction consistency between a perturbed student and an exponential-moving-average teacher. This is well suited to Sybil source-attacker traceability, where pseudonym switching, behavior manipulation, and dynamic topology may yield noisy pseudo-labels and unstable decision boundaries. Class imbalance is addressed by the
class-weighted focal loss.

The student model $f_\theta$ and teacher model $f_\phi$ share identical MLP backbones to map the spatio-temporal embedding $\mathbf{z}_i^{\mathrm{st}}$ to traceability class probabilities. To provide stable consistency targets, the teacher parameters $\phi$ are updated via an Exponentially Moving Average ($\mathrm{EMA}$) of the student parameters $\theta$ at each training iteration $k$.
\begin{equation}
\phi^{(k)} = \alpha \phi^{(k-1)} + (1-\alpha)\theta^{(k)}
\end{equation}
where $\alpha \in [0,1]$ is the $\mathrm{EMA}$ decay coefficient.

\textbf{Multi-level perturbation mechanisms.} 
A multi-level perturbation strategy is enforced to prevent representation collapse and guide the student network $f_\theta$ to learn noise-tolerant structural logic. At the graph level, an edge-dropping mask $\mathrm{drop}_{\mathrm{edge}}$ is applied within the student's prior $\mathrm{GNN}$ layers to perturb the dynamic topology. At the feature level, a masking matrix $\mathrm{mask}_{\mathrm{feat}}$ randomly injects corruptions into the intermediate features. Formally, using the comprehensive spatio-temporal embedding matrix $\mathbf{Z}_{\mathrm{st}}$ from $\mathrm{MSTA}$ as the base, the perturbed input embedding $\tilde{\mathbf{z}}_i^{\mathrm{st}}$ delivered to the student $\mathrm{MLP}$ is formulated as follows.
\begin{equation}
\tilde{\mathbf{z}}_i^{\mathrm{st}} = \psi(\mathbf{z}_i^{\mathrm{st}}, \mathrm{mask}_{\mathrm{feat}})
\end{equation}
where $\psi(\cdot)$ denotes the stochastic feature corruption operator.

\textbf{Training objective.}
The total objective balances the supervised loss
$\mathcal{L}_{\mathrm{sup}}$ and the consistency loss
$\mathcal{L}_{\mathrm{cons}}$:
\begin{subequations}\label{eq:joint_loss}
\begin{align}
\mathcal{L}_{\mathrm{total}}(\theta)&=\mathcal{L}_{\mathrm{sup}}(\theta)
+\lambda(k)\mathcal{L}_{\mathrm{cons}}(\theta),
\label{eq:total_loss}
\\
\mathcal{L}_{\mathrm{sup}}&=-\frac{1}{|\mathcal{D}_L|}\sum_{i\in\mathcal{D}_L}
\omega_i(1-p_{i,y})^\gamma
\log(p_{i,y}),
\label{eq:focal_loss}
\\
\mathcal{L}_{\mathrm{cons}}&=\frac{1}{|\mathcal{B}|}\sum_{i\in\mathcal{B}}
\mathbb{I}\left(\max\left(f_\phi(\mathbf{z}_i^{\mathrm{st}})\right)>\tau\right)
\nonumber\\
&\quad\times
\left\|
f_\theta(\tilde{\mathbf{z}}_i^{\mathrm{st}})-f_\phi(\mathbf{z}_i^{\mathrm{st}})
\right\|_2^2.
\label{eq:consistency_loss}
\end{align}
\end{subequations}

where $\mathcal{D}_L$ is the labeled set, $\mathcal{B}$ is an unlabeled mini-batch, $p_{i,y_i}$ is the student probability assigned to the ground-truth class $y_i$, $\omega_{y_i}$ is its class weight, and $\gamma$
is the focal parameter. The threshold $\tau$ retains only confident teacher predictions for consistency regularization.

To avoid noisy targets from destabilizing early training phases, the dynamic consistency weight $\lambda(k)$ follows a Gaussian ramp-up schedule over the training horizon:
\begin{equation}
\lambda(k) = \lambda_{\mathrm{max}} \exp\left(-5 \left(1-\frac{k}{K_{\mathrm{ramp}}}\right)^2\right)
\end{equation}
where $\lambda_{\mathrm{max}}$ represents the ceiling weight and $K_{\mathrm{ramp}}$ denotes the predefined duration of the ramp-up phase.

\section{Experiment}
\subsection{Experiment Setups}
\subsubsection{Datasets} 
The VeReMi-Extension dataset \cite{kamel2020MDS} was generated using the F2MD platform with Veins/OMNeT++ and SUMO under the LuSTNano traffic scenario. It contains vehicle mobility, V2X communication, pseudonyms, and binary benign/malicious annotations. In our work, simulator-only physical sender identifiers are used 
exclusively offline to associate pseudonyms with their physical sources and construct the three-class traceability labels. The main simulation and attack settings are summarized in Table~\ref{tab:simu_para}. 

\begin{figure}
    \begin{center}
        \includegraphics[width=3.5in]{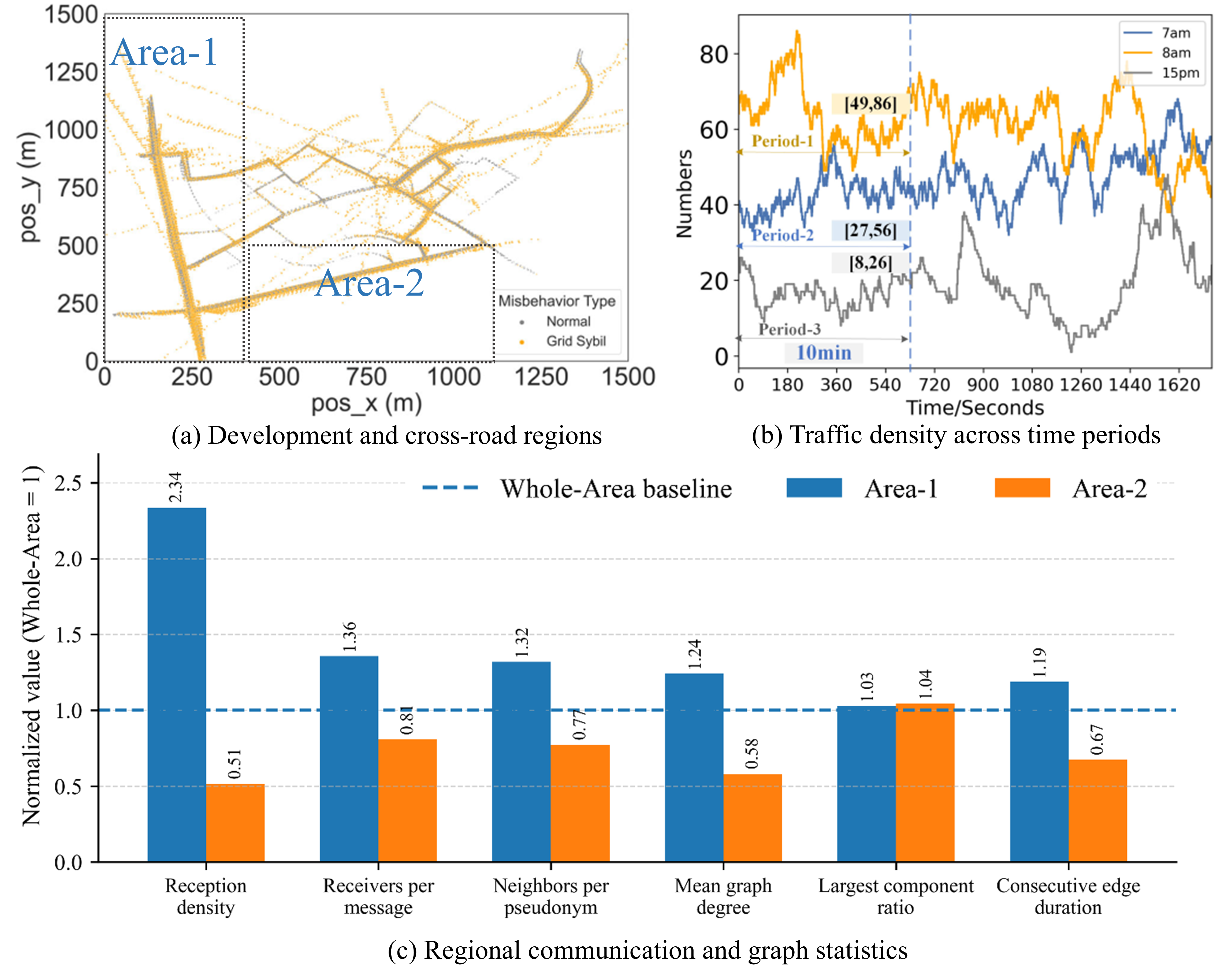}\\
        \caption{Simulation and generalization Settings}
        \label{fig:simulation_settings}
    \end{center}
\end{figure}

As illustrated in Fig.~\ref{fig:simulation_settings}, Area-1 corresponds to the Route d'Esch development region, whereas Area-2 provides a spatially disjoint cross-road setting. The temporal traffic profiles motivate the use of 08:00:00--08:09:59 as the high-density development
interval and 15:00:00--15:09:59 for cross-density evaluation, while the regional graph statistics confirm distinct communication-topology characteristics across the two areas. We therefore use the $1{,}500~\mathrm{m}\times400~\mathrm{m}$ Route d'Esch segment during
08:00:00--08:09:59 for model development and chronologically split the data into training, validation, and test sets at a ratio of 70:15:15. Model selection is performed exclusively on the validation set. The fixed model is subsequently evaluated on spatially disjoint road segments for cross-road generalization and on the same Route d'Esch segment during 15:00:00--15:09:59 for cross-density generalization. The processed datasets are publicly available\footnotemark{}. \footnotetext{https://github.com/octoberzzzzz/Four\_Sybil}


\begin{table}[htbp]
\centering
\caption{Dataset and evaluation settings.}
\label{tab:simu_para}
\begin{tabular}{ll}
\toprule
\textbf{Parameter} & \textbf{Value} \\
\midrule
Whole simulation area
& $1500\,\mathrm{m}\times1500\,\mathrm{m}$ \\
Development region& Route d'Esch, $400\,\mathrm{m}\times1500\,\mathrm{m}$ \\
Development interval& 08:00:00--08:09:59 \\
Cross-density interval& 15:00:00--15:09:59 \\
Cross-road scope& Route d'Esch, $[400,1250]\,\mathrm{m}\times[0,500]\,\mathrm{m}$ \\
Maximum vehicle speed& $50\,\mathrm{m/s}$ \\
Maximum acceleration& $3\,\mathrm{m/s^2}$ \\
Maximum deceleration& $5\,\mathrm{m/s^2}$ \\
Message interval
& $1\,\mathrm{s}$ \\
Pseudonym update interval
& $3\,\mathrm{min}$ \\
A1 records& 347,245 \\
A2 records& 179,520 \\
A3 records& 221,503 \\
A4 records& 175,721 \\
\bottomrule
\end{tabular}
\end{table}

\subsubsection{Evaluation Metrics}

\textbf{(1) Sybil attack pre-screening.}
Sybil attack pre-screening is formulated as a binary classification task. Accuracy and F1-score are adopted to evaluate its detection performance:

\begin{subequations}\label{eq:binary_metrics}
\begin{align}
\mathrm{Acc}&=\frac{TP+TN}{TP+TN+FP+FN},\\
\mathrm{F1}&=\frac{2PR}{P+R}.
\end{align}
\end{subequations}

where $TP$, $TN$, $FP$, and $FN$ denote the numbers of true-positive,true-negative, false-positive, and false-negative samples, respectively, while $P$ and $R$ denote precision and recall.

\textbf{(2) Source-attacker traceability.}
Source-attacker traceability is formulated as a three-class problem involving normal nodes, source attackers, and Sybil identities. Macro-F1 evaluates class-balanced performance, while Source-Attacker Recall (SAR) and Normal False-Accusation Rate (NFAR) assess source identification and false accusations, respectively. SAR is reported separately because strong performance on the Normal and Sybil classes may mask poor source-attacker recall in Macro-F1. NFAR measures the proportion of normal vehicles falsely classified as malicious, reflecting the safety cost of false alarms.

Let
$\mathcal{C}
=
\{\mathrm{normal},\mathrm{source},\mathrm{Sybil}\}$
denote the set of traceability classes and
$\mathcal{C}_{\mathrm{atk}}
=
\{\mathrm{source},\mathrm{Sybil}\}$
denote the set of malicious classes. The three metrics are defined as:

\begin{subequations}\label{eq:traceability_metrics}
\begin{align}
\mathrm{Macro\text{-}F1}
&=\frac{1}{|\mathcal{C}|}
\sum_{c\in\mathcal{C}}
\frac{2P_cR_c}{P_c+R_c},
\\
\mathrm{SAR}&=\frac{\sum_{i\in\mathcal{D}_{\mathrm{test}}}\mathbb{I}
\left(y_i=\mathrm{source}
\land \hat{y}_i=\mathrm{source}\right)}{
\sum_{i\in\mathcal{D}_{\mathrm{test}}}\mathbb{I}
\left(
y_i=\mathrm{source}\right)
},
\\
\mathrm{NFAR}
&=
\frac{
\sum_{i\in\mathcal{D}_{\mathrm{test}}}
\mathbb{I}
\left(
y_i=\mathrm{normal}\land\hat{y}_i\in\mathcal{C}_{\mathrm{atk}}\right)}{
\sum_{i\in\mathcal{D}_{\mathrm{test}}}\mathbb{I}
\left(
y_i=\mathrm{normal}\right)
}.
\end{align}
\end{subequations}

where $P_c$ and $R_c$ denote the precision and recall of class $c$, respectively; $y_i$ and $\hat{y}_i$ denote the ground-truth and predicted classes of test instance $i$; $\mathcal{D}_{\mathrm{test}}$ denotes the test set; and $\mathbb{I}(\cdot)$ is the indicator
function. SAR measures the proportion of source-attacker instances correctly identified as source attackers, whereas NFAR measures the proportion of normal instances incorrectly classified as either source attackers or Sybil identities.

\subsubsection{Experimental Details and Baselines}
All experiments are conducted on an NVIDIA GeForce RTX 5090 GPU with 32~GB VRAM. All deep learning models are developed using PyTorch. Semi-supervised baselines are implemented via the SemiLearn library, while online incremental streaming baselines are implemented using the River framework.

\textbf{(1) Baselines for Sybil Attack Pre-screening.} We evaluate three categories of pre-screening baselines: (i) \textit{Physical Layer Defenses}, including Threshold-based \cite{kamel2020MDS} and DTW \cite{sultana2024coop}; (ii) \textit{Tabular Supervised ML}, including GBDT \cite{Chen2022Sybil}, CNN-LSTM \cite{sultana2024Sybil}, VAN-IDS \cite{chen2024fast}, and Deep Ensemble TL \cite{shahid2025securing}; and (iii) \textit{Unsupervised and Incremental ML}, including VADGAN \cite{Devika2024VADGAN} and River-based incremental clustering modules \cite{river_doc_overview_2026} (KMeans, ODAC, CluStream, DBSTREAM, DenStream, STKMeans).

\textbf{(2) Baselines for Source Attacker Traceability.} All traceability baselines are independently re-implemented on a unified platform using SemiLearn under identical feature settings: (i) \textit{Tabular Supervised ML} (GBDT, XGBoost, Random Forest, CNN, BiLSTM); (ii) \textit{Tabular Semi-Supervised ML} combining paradigms (PseudoLabel, MeanTeacher, FreeMatch) with backbones (CNN, BiLSTM, Transformer); and (iii) \textit{Graph Semi-Supervised Models} (GCN-BiLSTM with MeanTeacher).

\subsection{Features Exploration}
\subsubsection{Spatial Distribution}
\begin{figure}[h]
    \vspace{-8pt}
    \centering
    \begin{subfigure}[b]{0.26\textwidth}
        \includegraphics[width=\textwidth]{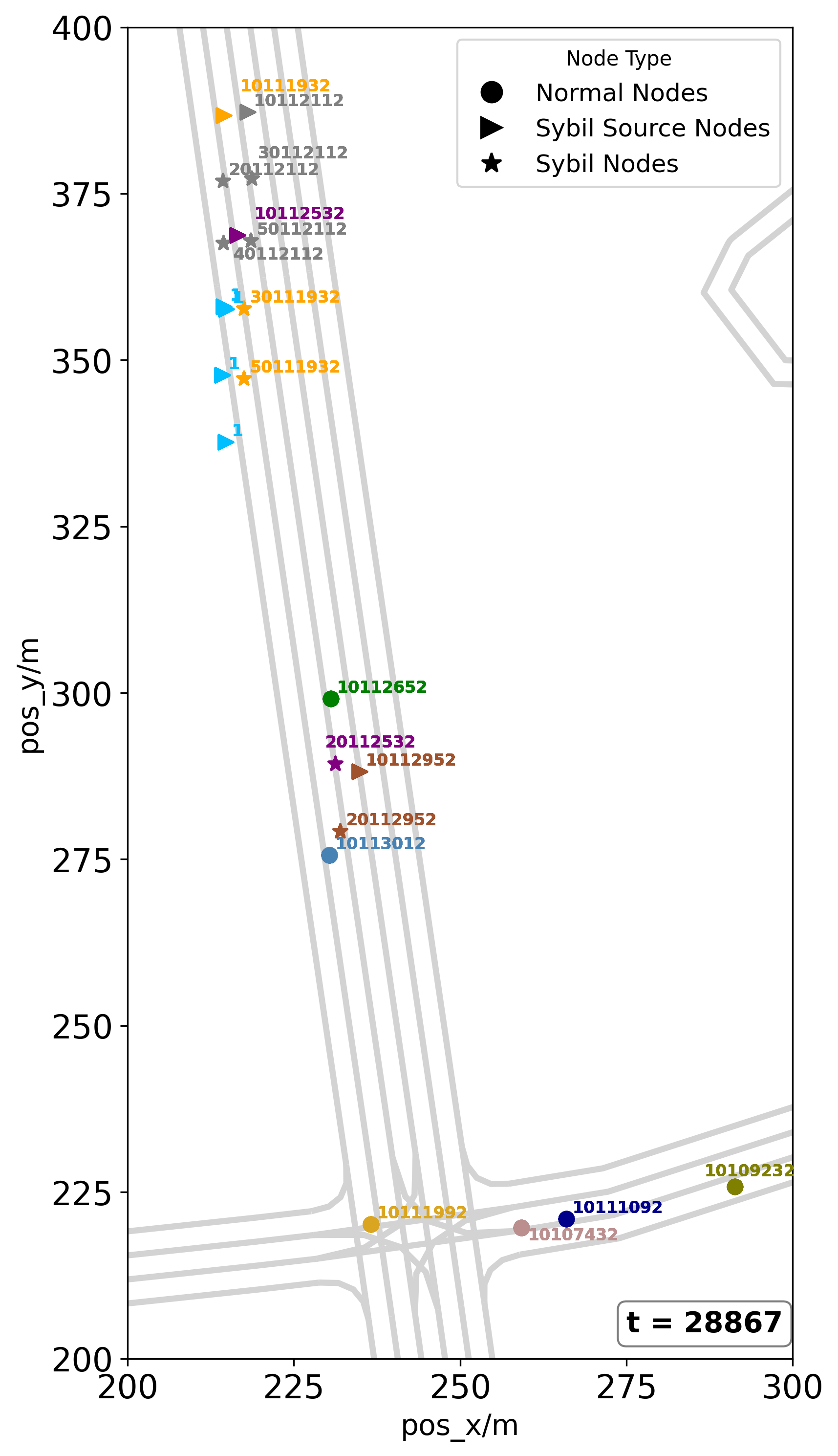}
        \caption{A1-Grid Sybil}
        \label{fig:A1_pos}
    \end{subfigure}
    \hspace{-0.1cm} 
    \begin{subfigure}[b]{0.22\textwidth}
        \includegraphics[width=\textwidth]{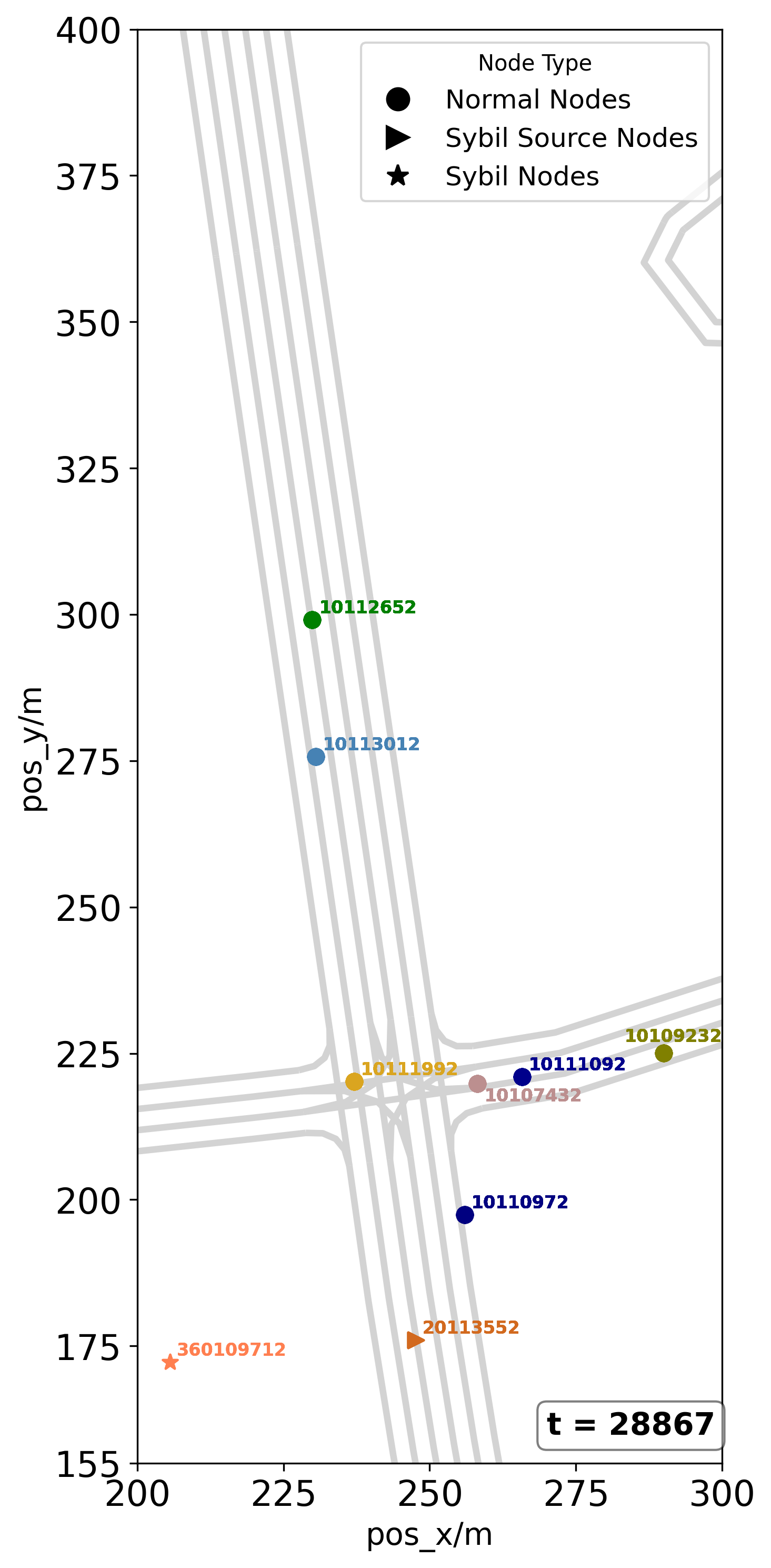}
        \caption{A2-DoS Random Sybil}
        \label{fig:A2_pos}
    \end{subfigure}
    \caption{Single-frame spatial distribution of A1 and A2, where the annotations are pseudonyms.}
    \label{fig:posfea}
    \vspace{-8pt}
\end{figure}

From a single-frame perspective, snapshot samples at $t = 28867$ were randomly selected. Single frames capture only the instantaneous spatial distribution of nodes, while differences between A2, A3, and A4 appear mainly in multiple frames and in the temporal domain. Hence, A2 serves as a representative for A3 and A4 due to their similar single-frame spatial characteristics. In Fig.\ref{fig:A1_pos}, the source attacker in A1 exploits legitimate pseudonyms to fabricate multiple virtual node clusters on the road, exhibiting an extremely regular geometric alignment.
This grid distribution contrasts sharply with the random vehicle distribution in normal traffic flow, exposing a clear position anomaly.
In Fig.~\ref{fig:A2_pos}, the source attacker in A2 broadcasts BSMs containing random kinematic parameters at an ultra-high frequency. 
These randomly generated positions cause attack nodes to appear as discontinuous, erratic position hops rather than a smooth trajectory.

\subsubsection{Temporal Distribution}
\begin{figure}[h]
    \vspace{-8pt}
    \centering
    \begin{subfigure}[b]{0.4\textwidth}
        \includegraphics[width=\textwidth]{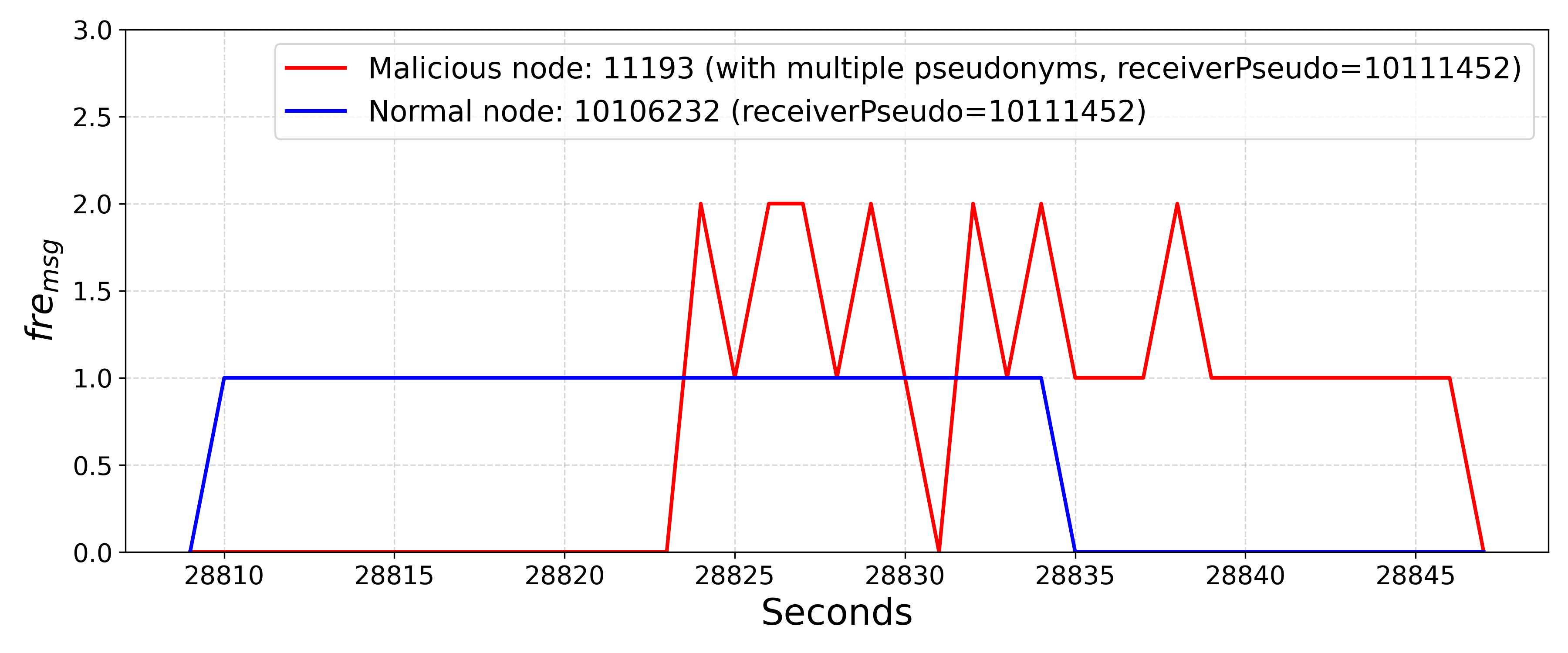}
        \caption{A3-DoS Disruptive Sybil}
        \label{fig:A3_temporal}
    \end{subfigure}
    \hspace{0.6cm} 
    \begin{subfigure}[b]{0.4\textwidth}
        \includegraphics[width=\textwidth]{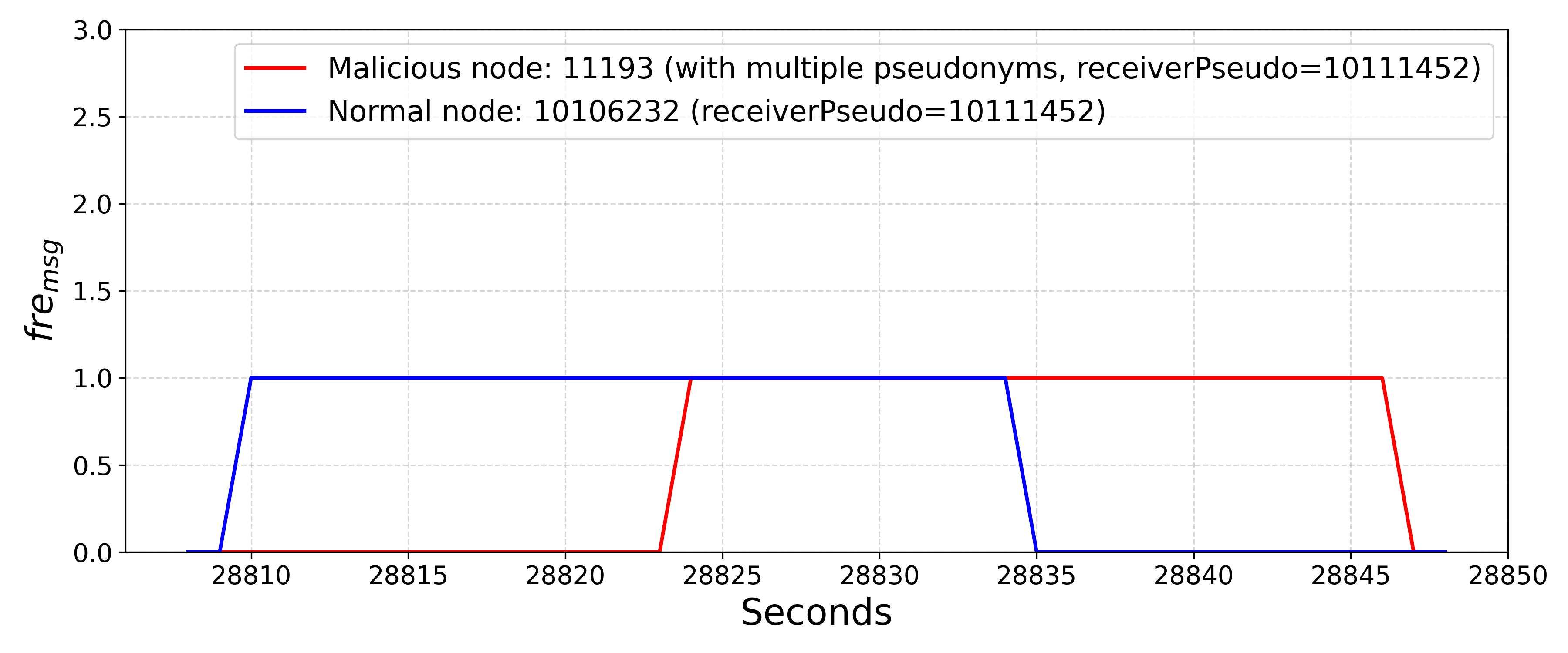}
        \caption{A4-Data Replay Sybil}
        \label{fig:A4_temporal}
    \end{subfigure}
    \caption{Temporal distribution of A3 and A4}
    \label{fig:posfea}
    \vspace{-8pt}
\end{figure}

The main attack characteristics of A1 have been revealed through single-frame spatial distribution analysis, while A2 shares similar DoS characteristics with A3. Therefore, A3 and A4 are further analyzed through temporal distribution to reveal differences in their dynamic communication behavior. Compared with kinematic features such as position and velocity, \(fre_{msg}\) can more directly reflect the temporal communication characteristics of Sybil attacks. In A3, malicious nodes exhibit significant fluctuations in message frequency in Fig.\ref{fig:A3_temporal}, showing bursty high-frequency communication behavior that disrupts normal communication patterns and increases channel congestion. In contrast, in A4, the message frequency of malicious nodes remains relatively similar to that of normal vehicles in Fig.\ref{fig:A4_temporal}, indicating that the attacker maintains normal temporal transmission patterns by replaying historical BSM sequences.

\begin{figure}[h]
    \vspace{-8pt}
    \centering

    \begin{subfigure}[b]{0.23\textwidth}
        \includegraphics[width=\textwidth]{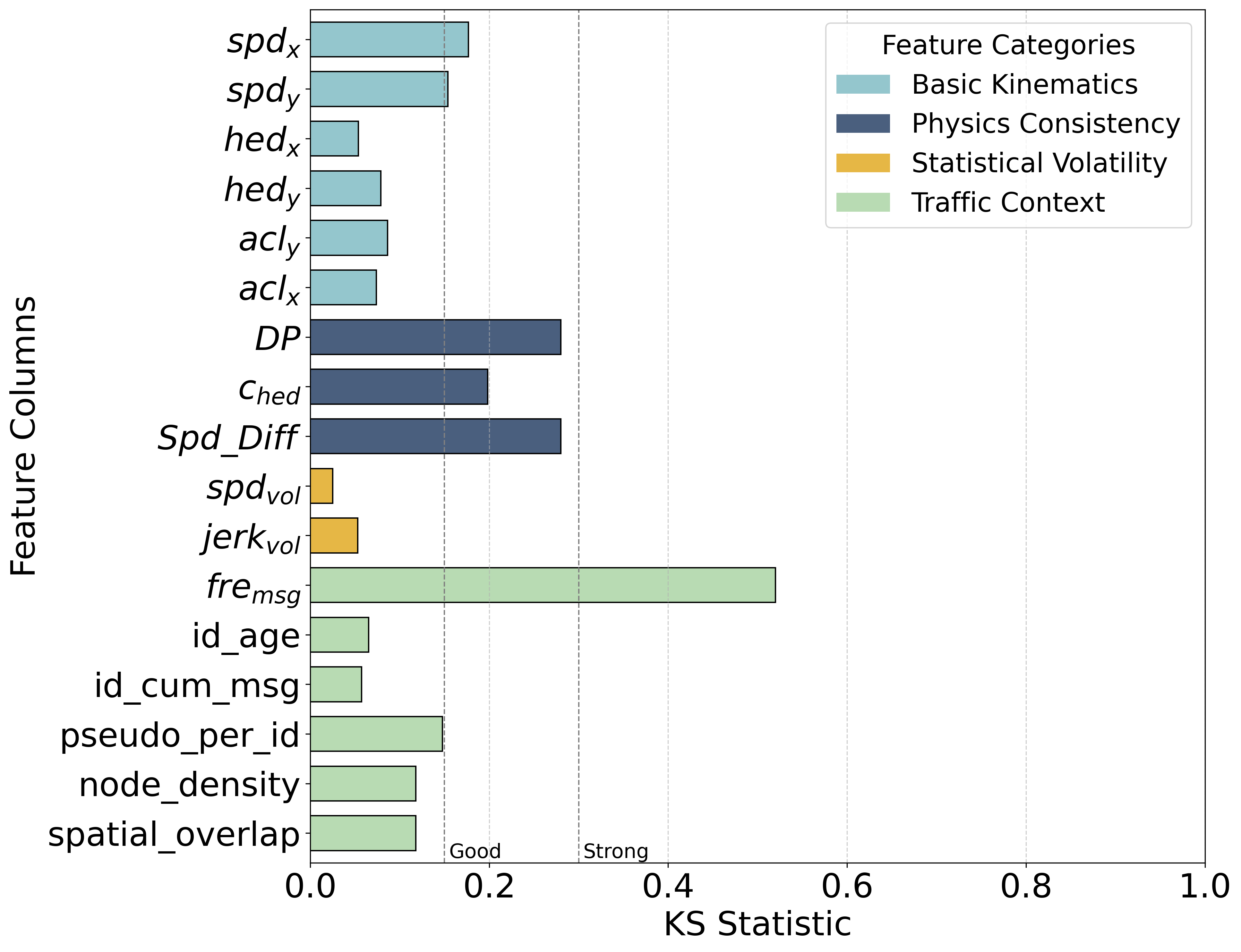}
        \caption{A1-Grid Sybil}
        \label{fig:A1_ks_node_attack}
    \end{subfigure}
    \hspace{-0.1cm}
    \begin{subfigure}[b]{0.23\textwidth}
        \includegraphics[width=\textwidth]{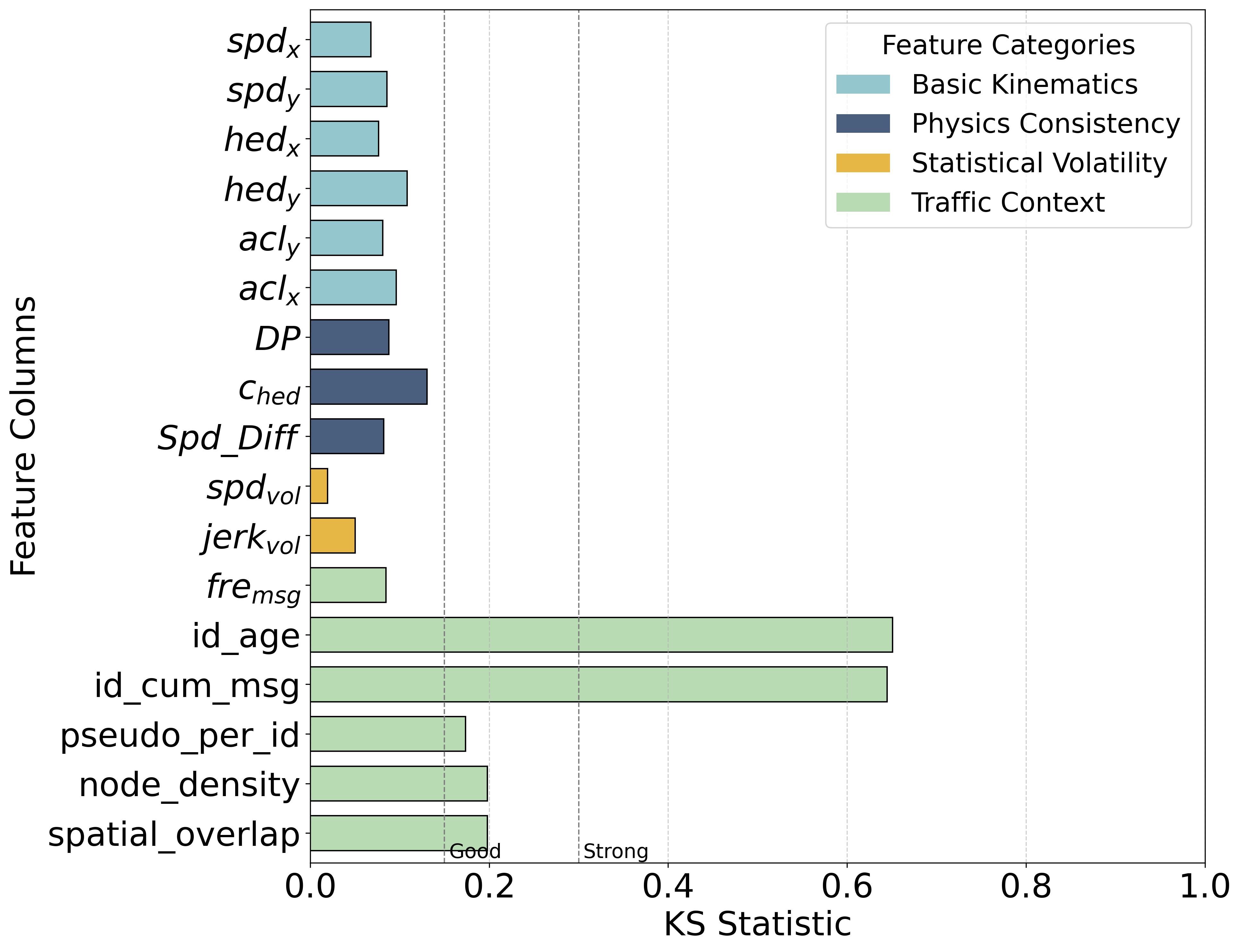}
        \caption{A2-DoS Random Sybil}
        \label{fig:A2_ks_node_attack}
    \end{subfigure}
    \hspace{-0.1cm}
    \begin{subfigure}[b]{0.23\textwidth}
        \includegraphics[width=\textwidth]{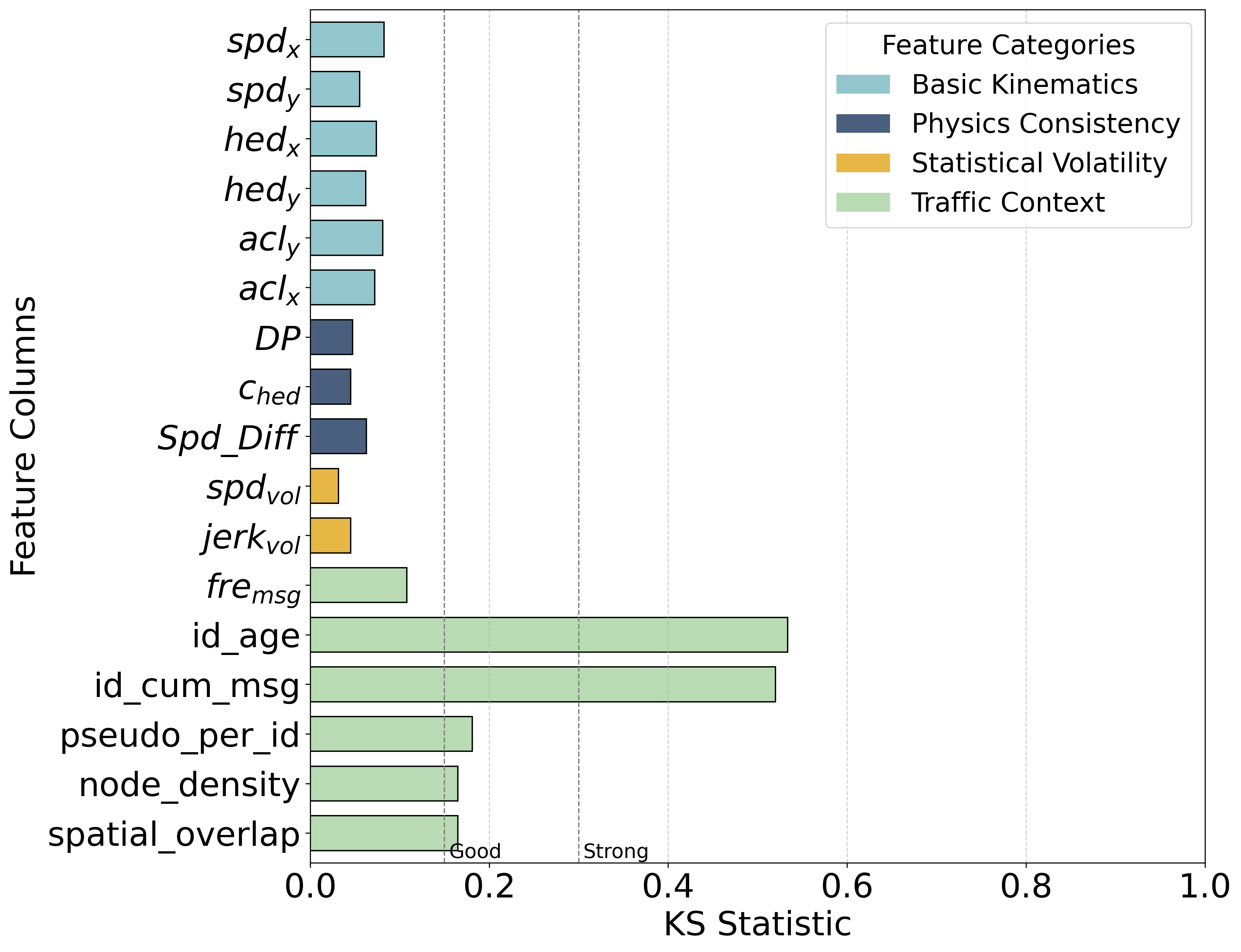}
        \caption{A3-DoS Disruptive Sybil}
        \label{fig:A3_ks_node_attack}
    \end{subfigure}
    \hspace{-0.1cm}
    \begin{subfigure}[b]{0.23\textwidth}
        \includegraphics[width=\textwidth]{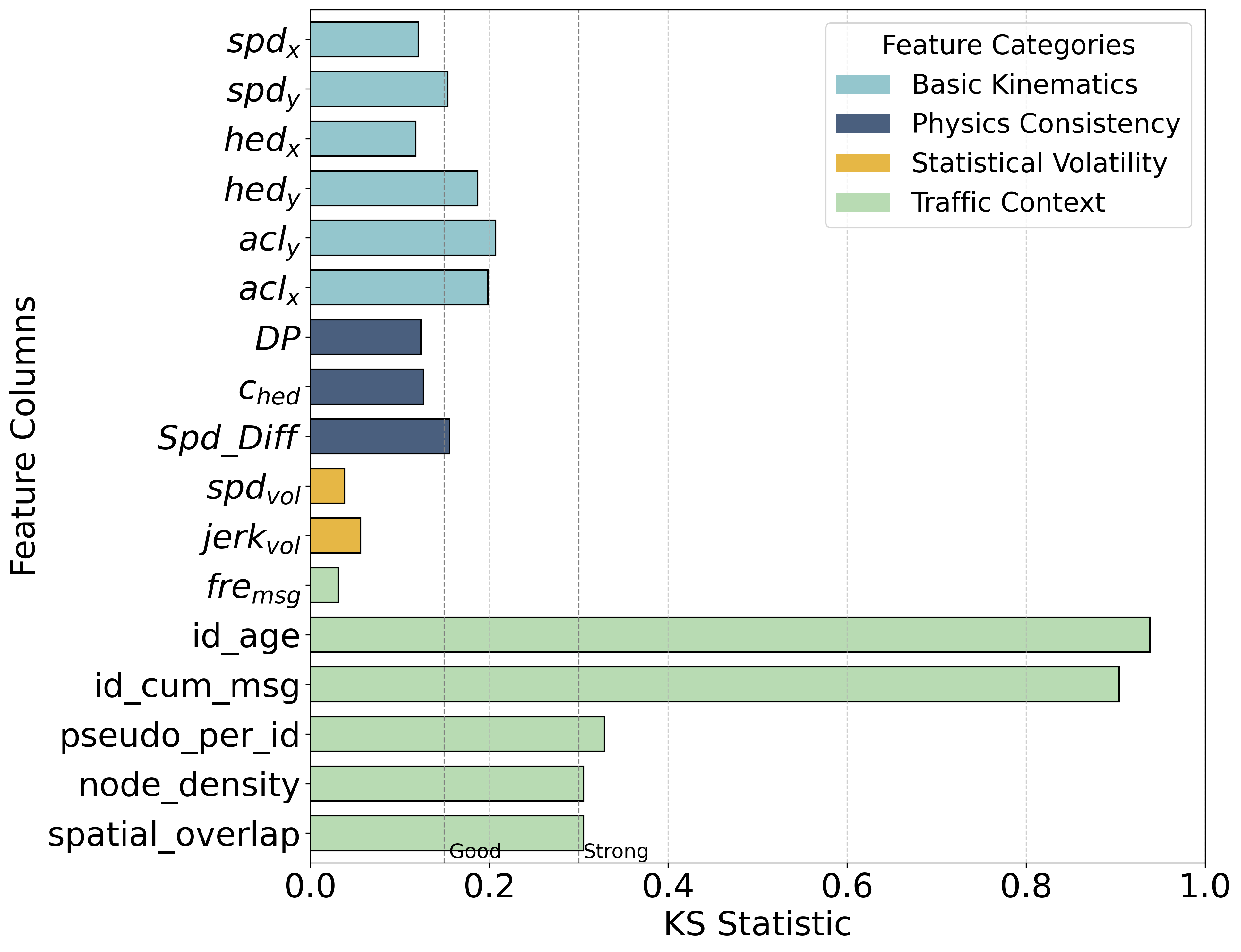}
        \caption{A4-Data Replay Sybil}
        \label{fig:A4_ks_node_attack}
    \end{subfigure}

    \caption{
    KS-based feature importance for source attacker traceability under four Sybil attack scenarios.
    }

    \label{fig:ks_traceability}

    \vspace{-10pt}

\end{figure}

\subsubsection{Feature Importance}
Results show that varying Sybil patterns yield distinct feature distribution separations, validating our lightweight KS-based screening strategy. In A1, traffic context (e.g., $fre_{msg}$) and physics consistency (e.g., $Spd\_Diff$) features exhibit high to moderate discriminability due to regular grid-structured behaviors. In A2, traffic context features dominate, particularly identity-lifecycle indicators (e.g., $id\_age$, $id\_cum\_msg$), due to abnormal node persistence and communication volumes. For A3 and A4, attackers mimic and replay legitimate trajectories, rendering traditional basic kinematics and physics consistency features ineffective. Nevertheless, traffic context features consistently yield the highest KS statistics. Overall, these results prove that exclusive reliance on basic kinematic features compromises Sybil traceability robustness, necessitating specific feature optimization.

\subsection{Qualitative Results of Source Attacker Traceability}

\begin{figure}
    \begin{center}
        \includegraphics[width=3.5in]{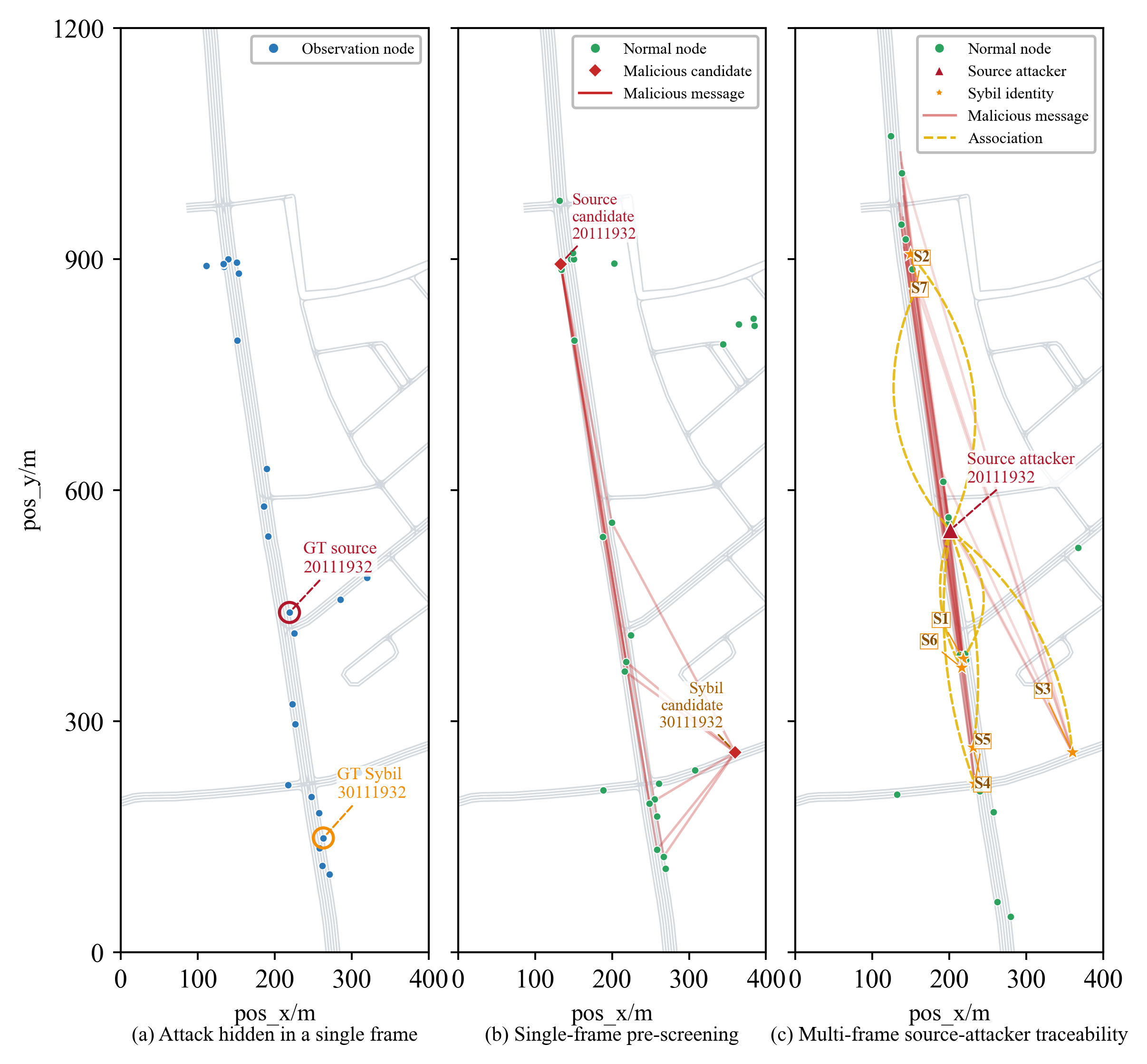}\\
        \caption{Qualitative results of multi-frame source-attacker traceability under the A3 DoS Disruptive Sybil attack.}
        \label{fig:qual}
    \end{center}
\end{figure}

We select the A3 DoS Disruptive Sybil attack as a representative case study to illustrate how Sybil-TraceGuard progressively traces rapidly changing Sybil identities back to their physical source attacker. As shown in Fig.~\ref{fig:qual}(a), the source pseudonym $20111932$ and its generated Sybil identity $30111932$ are already present in the scene. However, their malicious roles and source--Sybil relationship cannot be inferred from a single-frame spatial observation. Hence, all nodes are initially treated as observation nodes, while the highlighted circles are shown only as ground-truth references. In Fig.~\ref{fig:qual}(b), the incremental pre-screening stage flags the two pseudonyms as malicious candidates and highlights their suspicious message interactions. Although individual malicious observations may remain spatially plausible, the single-frame evidence is still insufficient to determine their common physical source. After aggregating observations over a five-second window, Fig.~\ref{fig:qual}(c) exploits the accumulated spatial, temporal, and communication dependencies to associate the short-lived Sybil identities $S1$--$S7$ with the same source attacker $20111932$. The yellow dashed curves represent inferred source--Sybil associations rather than physical vehicle trajectories. This example demonstrates that multi-frame cross-identity dependencies enable Sybil-TraceGuard to recover the common physical source behind rapidly changing Sybil pseudonyms.

\subsection{Quantitative Results on Sybil Attack Pre-Screening}

\begin{table*}[t]
\centering
\caption{Sybil attack pre-screening results (\%). Sup.: supervised; Unsup.\&Incre.: unsupervised and incremental. Gray shading denotes the best result within each category, while bold values denote the overall best result.}
\setlength{\tabcolsep}{1pt} 

\begin{footnotesize} 
\begin{sc}
\newcolumntype{C}{>{\centering\arraybackslash}X}
\label{tab:sybil_det}
\begin{tabularx}{\textwidth}{p{1.4cm} >{\raggedright\arraybackslash}p{3.1cm} CC CC CC CC} 
\toprule
\multirow{2}{*}{Type} & \multirow{2}{*}{Method} & \multicolumn{2}{c}{A1} & \multicolumn{2}{c}{A2} & \multicolumn{2}{c}{A3} & \multicolumn{2}{c}{A4} \\
\cmidrule(lr){3-4} \cmidrule(lr){5-6} \cmidrule(lr){7-8} \cmidrule(lr){9-10}
& & Acc $\uparrow$ & F1 $\uparrow$ & Acc $\uparrow$ & F1 $\uparrow$ & Acc $\uparrow$ & F1 $\uparrow$ & Acc $\uparrow$ & F1 $\uparrow$ \\

\midrule
\multirow{2}{1.4cm}{Physical} 
& DTW\cite{sultana2024coop}           & 98.61 & \colorbox{gray!20}{94.88} & N/A   & N/A & N/A & N/A & N/A & N/A \\
& Threshold-based\cite{kamel2020MDS}  & 82.04 & 74.15 & 98.98  & \colorbox{gray!20}{98.86} & 77.87 & \colorbox{gray!20}{69.14} & \colorbox{gray!20}{79.48} & 52.35 \\

\midrule

\multirow{5}{1.4cm}{Tabular Sup. ML} 
& GBDT\cite{Chen2022Sybil}           & 93.68 & 77.06 & 95.57 & 87.07 & N/A & N/A & N/A & N/A \\
& CNN-LSTM\cite{sultana2024Sybil}    & 96.58 & 92.81 & N/A & N/A & N/A & N/A & N/A & N/A \\
& VAN-IDS\cite{chen2024fast}         & \multicolumn{4}{c}{Mixed Dataset(A1+A2): Acc-99.55/F1-98.45} & N/A & N/A & N/A & N/A \\
& Deep Ensemble TL\cite{shahid2025securing} & 82.70 & 88.40 & 98.90 & \colorbox{gray!20}{99.90} & 79.50 & \colorbox{gray!20}{99.30} & \colorbox{gray!20}{85.80} & \colorbox{gray!20}{67.00} \\

\midrule

\multirow{1}{1.4cm}{Unsup.ML} 
& VADGAN\cite{Devika2024VADGAN}      & N/A & \colorbox{gray!20}{100.00} & N/A & \colorbox{gray!20}{93.33} & N/A & \colorbox{gray!20}{76.35} & N/A & \colorbox{gray!20}{88.50} \\
& KMeans        & 89.26 & 91.22 & 90.92 & 79.73 & 82.55 & 77.16 & \colorbox{gray!20}{85.36} & 71.22 \\
\midrule
\multirow{7}{1.4cm}{Unsup. \& Incre. ML}

& ODAC          & 87.94 & 90.09 & 76.10 & 38.93 & 85.27 & 68.91 & 76.42 & 53.66 \\
& CluStream     & 90.67 & 92.42 & 97.33 & 94.17 & 90.27 & 87.29 & 88.85 & 79.63 \\
& DBSTREAM      & \textbf{95.91} & \textbf{96.70} & 93.03 & 85.36 & 89.70 & 87.25 & 99.07 & 83.39 \\
& STKMeans      & 90.92 & 92.62 & 97.31 & 94.10 & 95.15 & 93.89 & 97.40 & 95.24 \\
& Ours(DenStream) & 94.58 & 95.60 & \textbf{99.36} & \textbf{98.61} & \textbf{98.12} & \textbf{96.56} & \textbf{97.77} & \textbf{97.23} \\
\bottomrule
\end{tabularx}
\end{sc}
\end{footnotesize}
\end{table*}

Existing studies rarely cover all four Sybil attacks simultaneously, 
except for \cite{kamel2020MDS} and \cite{shahid2025securing}, emphasizing the completeness of our evaluation. As shown in Table \ref{tab:sybil_det}, in terms of physical layer defense mechanisms, it exhibits a highly polarized performance across scenarios. The threshold-based method \cite{kamel2020MDS} achieves good performance under A2 with a $98.86\%$ F1-score; however, its detection capability severely degrades in A1, A3 and A4. Conversely, the DTW-based approach \cite{sultana2024coop} achieves a robust F1-score of $94.88\%$ under A1, while leaving the remaining scenarios unaddressed. In terms of tabular supervised ML models, it achieves competitive metrics but suffers from limited generalization. CNN-LSTM\cite{sultana2024Sybil} outperforms both GBDT \cite{Chen2022Sybil} and deep ensemble TL\cite{shahid2025securing} in A1. Meanwhile, VAN-IDS \cite{chen2024fast} achieves satisfactory performance only on a mixed dataset of A1 and A2. Deep Ensemble TL \cite{shahid2025securing} experiences a noticeable performance drop(Acc $79.50\%$ in A3 and F1-score $67.00\%$ in A4).

Regarding unsupervised baselines, VADGAN \cite{Devika2024VADGAN} delivers an insufficient F1-score of $76.35\%$ under A3, while KMeans yields degraded F1-scores across A2 ($79.73\%$), A3 ($77.16\%$), and A4 ($71.22\%$). Conversely, except for ODAC, the incremental streaming paradigm achieves robust and consistent results across all Sybil attacks. Although slightly lower than DBSTREAM in A1, our DenStream-based Sybil detection framework demonstrates obvious advantages in A2, A3, and A4. Overall, while remaining label-independent, DenStream establishes new benchmarks across A1–A4 with F1-scores of $95.60\%$, $98.61\%$, $96.56\%$, and $97.23\%$, respectively, outperforming the supervised solutions like deep ensemble TL \cite{shahid2025securing}.

\subsection{Quantitative Results on Source Attacker Traceability}


\newcommand{\metricthree}[3]{#1 & #2 & #3}

\begin{table*}[t]
\centering
\caption{Source-attacker traceability results (\%) with optimal $w$ and $r$ at $\rho=0.8$. M-F1, SAR, and NFAR denote Macro-F1, Source-Attacker Recall, and Normal False-Accusation Rate, respectively. Gray shading denotes the best result within each category, while bold values indicate the proposed Sybil-TraceGuard.}
\label{tab:sybil_trace}

\setlength{\tabcolsep}{2pt}
\renewcommand{\arraystretch}{1.05}

\resizebox{\textwidth}{!}{%
\begin{tabular}{lll*{4}{ccc}}
\toprule

\multirow{2}{*}{TYPE}
& \multirow{2}{*}{BASE}
& \multirow{2}{*}{SEMI-SUP. METHOD}
& \multicolumn{3}{c}{A1}
& \multicolumn{3}{c}{A2}
& \multicolumn{3}{c}{A3}
& \multicolumn{3}{c}{A4}
\\

\cmidrule(lr){4-6}
\cmidrule(lr){7-9}
\cmidrule(lr){10-12}
\cmidrule(lr){13-15}

& &
& M-F1$\uparrow$
& SAR$\uparrow$
& NFAR$\downarrow$

& M-F1$\uparrow$
& SAR$\uparrow$
& NFAR$\downarrow$

& M-F1$\uparrow$
& SAR$\uparrow$
& NFAR$\downarrow$

& M-F1$\uparrow$
& SAR$\uparrow$
& NFAR$\downarrow$
\\

\midrule

Tabular Sup. ML
& GBDT
& N/A
& \metricthree{82.67}{60.69}{7.91}
& \metricthree{\colorbox{gray!20}{96.70}}{82.40}{0.00}
& \metricthree{89.78}{89.64}{12.10}
& \metricthree{97.70}{90.95}{0.00}
\\

& XGBoost
& N/A
& \metricthree{83.68}{60.22}{4.98}
& \metricthree{93.55}{68.27}{0.00}
& \metricthree{90.04}{91.58}{12.27}
& \metricthree{97.39}{89.10}{0.00}
\\

& Random Forest
& N/A
& \metricthree{83.08}{60.94}{6.56}
& \metricthree{96.25}{85.07}{0.00}
& \metricthree{91.48}{97.07}{12.11}
& \metricthree{\colorbox{gray!20}{97.75}}{90.02}{0.00}
\\

& CNN
& N/A
& \metricthree{\colorbox{gray!20}{88.83}}{96.92}{18.50}
& \metricthree{93.44}{73.91}{3.12}
& \metricthree{\colorbox{gray!20}{97.13}}{96.20}{5.29}
& \metricthree{97.41}{96.36}{6.86}
\\

& BiLSTM
& N/A
& \metricthree{79.61}{96.61}{36.99}
& \metricthree{91.25}{65.13}{5.12}
& \metricthree{95.56}{94.74}{0.00}
& \metricthree{96.07}{93.18}{0.54}
\\

\midrule

Sup. GNN
& THGAT \cite{chen2025sybil}
& N/A
& \metricthree{\colorbox{gray!20}{87.04}}{N/A}{N/A}
& \metricthree{N/A}{N/A}{N/A}
& \metricthree{N/A}{N/A}{N/A}
& \metricthree{N/A}{N/A}{N/A}
\\

\midrule

\multirow{10}{*}{Tabular Semi. ML}
& XGBOD
& N/A
& \metricthree{82.31}{60.61}{8.53}
& \metricthree{92.24}{63.00}{0.00}
& \metricthree{89.55}{71.12}{5.37}
& \metricthree{87.25}{63.55}{0.00}
\\

\cmidrule(lr){2-15}

& CNN 
& PseudoLabel
& \metricthree{71.71}{26.28}{3.45}
& \metricthree{89.38}{62.50}{2.40}
& \metricthree{\colorbox{gray!20}{96.80}}{84.92}{0.00}
& \metricthree{87.39}{59.39}{14.22}
\\

&
& FreeMatch
& \metricthree{62.04}{16.13}{9.75}
& \metricthree{96.48}{86.21}{0.00}
& \metricthree{95.91}{89.66}{2.55}
& \metricthree{92.70}{85.06}{0.44}
\\

&
& MeanTeacher
& \metricthree{63.72}{47.29}{38.06}
& \metricthree{84.10}{51.74}{8.39}
& \metricthree{92.74}{94.64}{20.53}
& \metricthree{90.36}{88.24}{6.35}
\\

\cmidrule(lr){2-15}

& BiLSTM
& PseudoLabel
& \metricthree{\colorbox{gray!20}{93.70}}{76.06}{0.24}
& \metricthree{90.08}{89.55}{29.70}
& \metricthree{84.20}{64.86}{18.97}
& \metricthree{94.75}{80.00}{0.45}
\\

&
& FreeMatch
& \metricthree{61.24}{41.67}{23.50}
& \metricthree{96.70}{82.95}{0.00}
& \metricthree{87.13}{69.05}{0.00}
& \metricthree{95.70}{88.89}{0.14}
\\

&
& MeanTeacher
& \metricthree{73.74}{77.54}{41.42}
& \metricthree{93.25}{98.21}{20.98}
& \metricthree{93.83}{76.60}{3.97}
& \metricthree{95.57}{94.55}{6.10}
\\

\cmidrule(lr){2-15}

& Transformer
& PseudoLabel
& \metricthree{93.39}{84.56}{3.47}
& \metricthree{88.21}{79.10}{30.00}
& \metricthree{83.04}{44.68}{13.25}
& \metricthree{87.61}{53.33}{0.91}
\\

&
& FreeMatch
& \metricthree{63.49}{9.17}{8.56}
& \metricthree{94.16}{73.43}{0.00}
& \metricthree{93.77}{93.10}{0.00}
& \metricthree{96.62}{97.67}{0.22}
\\

&
& MeanTeacher
& \metricthree{80.82}{48.33}{9.54}
& \metricthree{93.05}{89.29}{16.03}
& \metricthree{94.19}{85.19}{0.17}
& \metricthree{\colorbox{gray!20}{98.15}}{91.43}{1.05}
\\

\midrule

Graph Semi. ML
& GAT
& MeanTeacher
& \metricthree{80.02}{48.33}{9.52}
& \metricthree{86.33}{50.00}{5.78}
& \metricthree{77.01}{24.80}{8.11}
& \metricthree{92.62}{87.28}{1.51} 
\\

& GCN-BiLSTM
& MeanTeacher
& \metricthree{91.95}{86.36}{22.00}
& \metricthree{93.95}{76.92}{0.34}
& \metricthree{93.62}{70.06}{0.27}
& \metricthree{95.76}{78.56}{0.16}
\\

& \textbf{Sybil-TraceGuard}
& \textbf{MeanTeacher}
& \metricthree
  {\textbf{94.91}}
  {\textbf{96.88}}
  {\textbf{0.16}}
& \metricthree
  {\textbf{97.03}}
  {\textbf{95.45}}
  {\textbf{0.07}}
& \metricthree
  {\textbf{98.96}}
  {\textbf{99.16}}
  {\textbf{0.17}}
& \metricthree
  {\textbf{98.89}}
  {\textbf{100.00}}
  {\textbf{0.19}}
\\

\bottomrule
\end{tabular}}
\end{table*}

Several studies \cite{sultana2024coop,zhu2024sybil,chen2025sybil} have investigated Sybil detection or source-oriented traceability. However, \cite{sultana2024coop} and \cite{zhu2024sybil} are excluded from Table~\ref{tab:sybil_trace} because their evaluation protocols and reported metrics are not directly comparable with our three-class traceability setting. THGAT \cite{chen2025sybil} is retained where a
comparable result is available.

Overall, Sybil-TraceGuard achieves Macro-F1 scores of 94.91\%, 97.03\%, 98.96\%, and 98.89\% in A1--A4, respectively, outperforming the strongest competing methods in all four scenarios. The gain is particularly evident in A3 and A4, where it surpasses CNN (97.13\%) and Transformer with MeanTeacher (98.15\%), respectively. These results demonstrate consistently high class-balanced traceability performance across heterogeneous Sybil behaviors.

The SAR and NFAR results further show that this improvement is not achieved by simply increasing sensitivity to malicious nodes. Sybil-TraceGuard obtains SAR values of 96.88\%, 95.45\%, 99.16\%, and 100.00\% while keeping NFAR below 0.2\% in all four scenarios. In contrast, several competing methods achieve high SAR at the cost of substantially higher false accusations; for example, CNN reaches 96.92\% SAR in A1 with an NFAR of 18.50\%, while BiLSTM with MeanTeacher reaches 98.21\% SAR in A2 with an
NFAR of 20.98\%. This indicates that Sybil-TraceGuard provides a more favorable trade-off between source-attacker identification and protection of normal vehicles.

The results also show that the optimal semi-supervised strategy is attack- and backbone-dependent. PseudoLabel performs best for all three deep backbones in A1, whereas FreeMatch is consistently stronger in A2. For A3 and A4, the preferred strategy varies across backbones. This variation suggests that no generic SSL method is consistently optimal across heterogeneous Sybil patterns. In contrast, Sybil-TraceGuard combines spatial, communication, and temporal evidence within a unified MeanTeacher-based framework, resulting in more stable performance across the four attack scenarios.

\subsection{Sensitivity Analysis of Hyperparameters}

\subsubsection{window size ($w$) and step ratio ($r$)}
A1 serves as the primary case study since it is the sole scenario with a performance under 95\%. The Transformer with PseudoLabel serves to evaluate the necessity of spatial graph topology for source traceability. Meanwhile, GCN-BiLSTM with MeanTeacher provides a direct architectural comparison between standard graph baselines and our proposed Sybil-TraceGuard under identical consistency regularization. 
In Table \ref{tab:sybil_wr}, all models exhibit performance degradation as $r$ increases from 0.1,0.5,0.8 to 1.0, which can be attributed to the reduced temporal overlap and sample density. Notably, the GCN-BiLSTM baseline collapses to 40.12\% at $w=15, r=1.0$. Our method maintains robust and stable performance across all configurations, consistently exceeding 89.00\%. Furthermore, owing to the MSTA module, our model demonstrates remarkable scale invariance with marginal fluctuations under variations in $w$, confirming its effectiveness in capturing robust spatiotemporal features against diverse sampling frequencies.

\begin{table}[htbp]
\centering
\caption{Macro-F1 performance under different configurations of $w$ and $r$ in A1}
\label{tab:sybil_wr}
\renewcommand{\arraystretch}{1.3}
\begin{tabular}{@{}llcccc@{}}
\toprule

\textbf{Model} & \makecell[lb]{\textbf{Window}\\\textbf{Size ($|W|$)}} & \multicolumn{4}{c}{\textbf{Step Ratio ($r$)}} \\ \cmidrule(l){3-6}
 & & \textbf{0.1} & \textbf{0.5} & \textbf{0.8} & \textbf{1.0} \\ \midrule

\multirow{3}{*}{\makecell[l]{Transformer\\(PseudoLabel)}} 
 & 5  & 91.36 & 88.55 & 87.35 & 83.72 \\
 & 10 & 90.47 & 85.42 & 81.45 & 75.65 \\ 
 & 15 & \textbf{93.39} & 89.65 & 91.15 & 85.27 \\ \midrule

\multirow{3}{*}{\makecell[l]{GCN-BiLSTM\\(MeanTeacher)}} 
 & 5  & \textbf{91.95} & 90.91 & 90.54 & 87.65 \\
 & 10 & 91.71 & 89.89 & 87.68 & 80.87 \\ 
 & 15 & 91.72 & 83.35 & 67.19 & 40.12 \\ \bottomrule

\multirow{3}{*}{\makecell[l]{Sybil-TraceGuard}}       
 & 5  & \textbf{94.91} & 93.34 & 91.48 & 92.87 \\
 & 10 & 94.26 & 93.78 & 91.77 & 89.65 \\ 
 & 15 & 94.82 & 93.88 & 92.55 & 90.06 \\ \bottomrule
 
\end{tabular}
\end{table}

\subsubsection{Unlabeled ratio $\rho$}
In Table \ref{tab:unlabeled_ratio}, the settings of $w$ and $r$ are consistent with Table V. XGBOD is used to demonstrate the performance upper bound of traditional tabular classification via automated feature engineering, and BiLSTM (MeanTeacher) alongside GCN-BiLSTM (MeanTeacher) represent the SOTA in tabular and graph semi-supervised learning, respectively.
In Table \ref{tab:unlabeled_ratio}, our method outperforms all other methods across all $\rho$ values, peaking at a 95.30\% Macro-F1 score when $\rho=0.60$.Although performance predictably declines as $\rho$ increases due to diminishing supervision, our model remains remarkably stable, sustaining a marginal decay of merely 1.2\% even at the extreme $\rho=0.95$.
Conversely, XGBOD fluctuates significantly under high label scarcity.
Furthermore, while PseudoLabel-based Transformer and BiLSTM are competitive when labels are abundant, they are highly sensitive to label sparsity compared to MeanTeacher-based models, i.e., GCN-BiLSTM and ours. This confirms that combining Mean-Teacher with our spatio-temporal auditing logic effectively exploits unlabeled V2X streams, ensuring high traceability under limited supervision.

\begin{table}[htbp]
\centering
\caption{Macro-F1 performance under different $\rho$ in A1}
\label{tab:unlabeled_ratio}
\renewcommand{\arraystretch}{1.3}
\setlength{\tabcolsep}{6pt} 
\begin{tabular}{@{} >{\raggedright\arraybackslash}p{2cm} >{\centering\arraybackslash}p{1cm} ccccc @{}}
\toprule
\multirow{2}{3cm}{\textbf{Model}} & \multirow{2}{0.8cm}{\centering \makecell[c]{\textbf{$|W|,r$}}} & \multicolumn{5}{c}{\textbf{Unlabeled Ratio ($\rho$)}} \\ 
\cmidrule(l){3-7} 
 & & \textbf{0.60} & \textbf{0.70} & \textbf{0.80} & \textbf{0.90} & \textbf{0.95} \\ \midrule

XGBOD & N/A & 81.46 & 82.14 & 82.31 & 82.29 & 83.78 \\ \addlinespace[2pt]

\makecell[l]{Transformer\\(PseudoLabel)} & \makecell[c]{$|W|$=15\\r=0.1} & 93.70 & 93.34 & 93.39 & 91.82 & 90.97 \\ \addlinespace[2pt]

\makecell[l]{BiLSTM\\(PseudoLabel)} & \makecell[c]{$|W|$=10\\r=0.1} & 94.57 & 94.68 & 93.70 & 93.31 & 92.39 \\ \addlinespace[2pt]

\makecell[l]{GCN-BiLSTM\\(MeanTeacher)} & \makecell[c]{$|W|$=5\\r=0.1} & 92.11 & 92.05 & 91.95 & 91.51 & 91.46 \\ \addlinespace[2pt]

\makecell[l]{Sybil-TraceGuard} & \makecell[c]{$|W|$=5\\r=0.1} & 95.30 & 95.05 & 94.91 & 94.18 & 94.10 \\
\bottomrule
\end{tabular}
\end{table}

\subsection{Quantitative Results on Cross Road and Cross Density Generation}
Table~\ref{tab:cross_generalization} evaluates model generalization under spatial and traffic-density shifts using fixed models trained on the base scenario. Sybil-TraceGuard exhibits the most stable overall performance, achieving Macro-F1 scores of 93.09\% and 93.81\% under cross-road and cross-density settings, corresponding to only 0.82\% and 1.10\% relative drops, respectively. More importantly, it preserves
high SAR values of 94.13\% and 95.25\% while keeping NFAR below 0.2\% in both settings. In contrast, competing methods show larger performance degradation or a less favorable SAR--NFAR trade-off,
particularly under the cross-density shift. For example, XGBOD suffers a 15.48\% Macro-F1 drop with SAR decreasing to 24.18\%, while GCN-BiLSTM retains relatively high SAR but incurs NFARs of 23.95\%
and 18.93\%. These results demonstrate that Sybil-TraceGuard generalizes more reliably to unseen road segments and varying traffic densities without substantially increasing false accusations.

\newcommand{\mfdrop}[2]{%
#1\,{\scriptsize$_{\scriptscriptstyle(-#2\%)}$}%
}

\begin{table}[t]
\centering
\caption{Source-attacker traceability performance (\%) under
cross-road and cross-density generalization settings at $\rho=0.8$.
Small subscripts denote the relative M-F1 drop from the corresponding
base scenario.}
\label{tab:cross_generalization}

\renewcommand{\arraystretch}{1.25}
\setlength{\tabcolsep}{2.0pt}

\begin{tabular}{@{}
>{\raggedright\arraybackslash}p{1.55cm}
ccc
ccc
@{}}
\toprule

\multirow{2}{*}{\textbf{Model}}
& \multicolumn{3}{c}{\textbf{Cross-Road}}
& \multicolumn{3}{c}{\textbf{Cross-Density}}
\\

\cmidrule(lr){2-4}
\cmidrule(lr){5-7}

& \textbf{M-F1$\uparrow$}
& \textbf{SAR$\uparrow$}
& \textbf{NFAR$\downarrow$}
& \textbf{M-F1$\uparrow$}
& \textbf{SAR$\uparrow$}
& \textbf{NFAR$\downarrow$}
\\

\midrule

XGBOD
& \mfdrop{73.38}{8.93}& 62.87 & 11.37& \mfdrop{66.83}{15.48}& 24.18 & 11.44
\\
\addlinespace[2pt]

\makecell[l]{Transformer\\(PseudoLabel)}& \mfdrop{92.52}{0.87}& 76.51 & 1.85
& \mfdrop{91.33}{2.06}& 86.98 & 7.91
\\
\addlinespace[2pt]

\makecell[l]{BiLSTM\\(PseudoLabel)}
& \mfdrop{92.92}{0.78}& 76.26 & 2.08& \mfdrop{88.51}{5.19}& 60.08 & 0.72
\\
\addlinespace[2pt]

\makecell[l]{GCN-BiLSTM\\(MeanTeacher)}
& \mfdrop{88.75}{3.20}& 85.23 & 23.95& \mfdrop{85.90}{6.05}& 84.59 & 18.93
\\
\addlinespace[2pt]

\textbf{\makecell[l]{Sybil-\\TraceGuard}}
& \textbf{\mfdrop{93.09}{0.82}}& \textbf{94.13}& \textbf{0.11}& \textbf{\mfdrop{93.81}{1.1}}
& \textbf{95.25}& \textbf{0.15}
\\

\bottomrule
\end{tabular}
\end{table}

\subsection{Complexity Analysis}
Let $M$ denote the number of incoming BSM observations, $C_m$ the number of active micro-clusters, $N$ the number of pseudonym nodes forwarded to Stage II, $E$ the number of graph edges, $W$ the temporal sequence length, and $H$ the hidden dimension. ISAD processes the stream incrementally with a complexity of $\mathcal{O}(MC_md)$. For Stage II, DTC requires $\mathcal{O}(N^2+E\log E)$ due to pairwise spatial relation construction and edge deduplication; SGEM requires $\mathcal{O}(NWH^2+WEH)$; and MSTA requires
$\mathcal{O}[L_TN(WH^2+W^2H)]$. The node-wise AT classifier introduces only $\mathcal{O}(NH^2)$ complexity. Table~\ref{tab:complexity} compares the asymptotic inference complexity with representative deep-learning baselines used in our experiments.

\begin{table}[t]
\centering
\caption{Inference complexity of representative models.}
\label{tab:complexity}
\renewcommand{\arraystretch}{1.18}
\setlength{\tabcolsep}{4pt}

\begin{tabular}{@{}
>{\raggedright\arraybackslash}p{1.45cm}
>{\centering\arraybackslash}p{5.2cm}
@{}}
\toprule
\textbf{Model} & \textbf{Time Complexity} \\
\midrule

CNN& $\mathcal{O}(NWkH^2)$ \\

BiLSTM& $\mathcal{O}(NWH^2)$ \\

Transformer& $\mathcal{O}\!\left[L_TN(WH^2+W^2H)\right]$ \\

GCN-BiLSTM& \makecell[c]{$\mathcal{O}\!\left[WL_G(NH^2+EH)\right.$\\$\left.{}+NWH^2\right]$} \\

Sybil-TraceGuard
& \makecell[c]{
$\mathcal{O}\!\left(MC_md+N^2+E\log E+WEH\right.$\\
$\left.{}+L_TNW(H^2+WH)\right)$
} \\

\bottomrule
\end{tabular}
\end{table}

Although Sybil-TraceGuard introduces additional topology construction and multi-scale temporal auditing, ISAD filters the high-volume BSM stream before graph inference, such that Stage II operates only on a
reduced suspicious-node set ($N \ll M$). Moreover, the communication graph remains sparse and $W$ is bounded in our implementation. Mean-Teacher only adds a constant-factor training overhead and does not
increase online inference complexity.

\subsection{Ablation Experiment}
The module-level ablation results confirm the complementary roles of ISAD, DTC, SGEM, and MSTA. Removing ISAD causes larger degradation in A2 and A4, with the Macro-F1 of A2 dropping from 97.03\% to 85.43\%,
indicating the importance of incremental anomaly pre-screening for highly dynamic attack behaviors. Removing MSTA produces the most pronounced degradation in A1--A3; for example, the Macro-F1 in A2 decreases to 69.89\%. This result demonstrates that temporal auditing is critical for capturing communication-frequency variations and multi-frame inconsistencies that cannot be sufficiently characterized by spatial and topological relations alone. The performance drops caused by removing DTC or SGEM further confirm the importance of dynamic relation modeling and spatial-communication dependencies for
source-attacker traceability.

The learning-strategy ablations show that MeanTeacher, data perturbation, and Focal Loss are particularly important in A2 and A3, where attack behaviors exhibit stronger randomness and class ambiguity.
Removing MeanTeacher eliminates teacher-based consistency regularization, resulting in substantial degradation under these dynamic scenarios. Likewise, removing data perturbation weakens robustness to representation variations, while replacing Focal Loss with standard cross-entropy reduces the model's emphasis on difficult and minority samples. Together, these results show that the proposed learning strategy improves the robustness of Sybil-TraceGuard under heterogeneous and imbalanced attack patterns.

\begin{table}[htbp]
\centering
\caption{Ablation study results for Macro-F1 scores under four Sybil attack scenarios}
\label{tab:ablation}

\fontsize{8pt}{8pt}\selectfont
\renewcommand{\arraystretch}{1.15}
\setlength{\tabcolsep}{4.5pt}

\begin{tabular}{@{}lcccc@{}}
\toprule
\textbf{Config.}
& \textbf{A1}
& \textbf{A2}
& \textbf{A3}
& \textbf{A4}
\\
\midrule

\textbf{Full Model}
& \textbf{94.91}
& \textbf{97.03}
& \textbf{98.96}
& \textbf{98.89}
\\
\midrule

\rowcolor[HTML]{F2F2F2}
\textit{Modules}
& & & & \\
w/o ISAD
& 89.53 & 85.43 & 97.51 & 89.01 \\
w/o DTC
& 90.32 & 90.69 & 78.24 & 93.21 \\
w/o SGEM
& 89.92 & 88.03 & 78.50 & 92.97 \\
w/o MSTA
& 76.84 & 69.89 & 71.01 & 92.92 \\

\midrule

\rowcolor[HTML]{F2F2F2}
\textit{Learning Strategy}
& & & & \\
w/o Mean Teacher
& 91.53 & 87.41 & 74.34 & 93.46 \\
w/o Data Perturbation
& 92.26 & 71.92 & 82.10 & 91.17 \\

\midrule

\rowcolor[HTML]{F2F2F2}
\textit{Loss Function}
& & & & \\
w/o Focal Loss
& 91.91 & 68.78 & 89.65 & 93.06 \\

\bottomrule
\end{tabular}
\end{table}
    
\section{Conclusion}
Sybil-TraceGuard is a dynamic semi-supervised GNN framework that shifts Sybil defense from identity-level detection toward physical-source traceability. 
By integrating incremental anomaly pre-screening, dynamic topology construction, spatial-communication
modeling, and multi-scale temporal auditing, Sybil-TraceGuard associates fragmented and rapidly changing Sybil pseudonyms with their underlying source attackers under limited supervision. 
Experiments across four Sybil attack scenarios demonstrate consistently strong traceability
performance, achieving Macro-F1 scores of 94.91--98.96\%, Source-Attacker
Recall(SAR) of 95.45--100.00\%, and a Normal False-Accusation Rate (NFAR) below 0.2\%. 
Sensitivity, generalization, and ablation studies further confirm the robustness and complementary contributions of the proposed components and semi-supervised learning strategy.
Future work will extend the evaluation to real-world V2X traces and investigate adaptation to more complex Sybil behaviors under heterogeneous mobility and communication conditions.

\bibliographystyle{IEEEtran}
\bibliography{IEEEabrv,Bibliography}

@article{matin2022impacts,
  title={Impacts of connected and automated vehicles on road safety and efficiency: A systematic literature review},
  author={Matin, Ali and Dia, Hussein},
  journal=IEEE_TITS,
  volume={24},
  number={3},
  pages={2705--2736},
  year={2022},
  publisher={IEEE}
}

@article{pan2024impacts,
  title={The impacts of connected autonomous vehicles on mixed traffic flow: A comprehensive review},
  author={Pan, Yuchen and Wu, Yu and Xu, Lu and Xia, Chengyi and Olson, David L},
  journal={Physica A: Statistical Mechanics and its Applications},
  volume={635},
  pages={129454},
  year={2024},
  publisher={Elsevier}
}

@article{hua2026envir,
  title={Environmental sustainability and mobility impacts of connected and autonomous vehicles at urban highways},
  author={Hua, Chengying and Fan, Wei David and Hua, Wei and Zhang, Shuichao},
  journal={Expert Syst. Appl.},
  volume={295},
  pages={128892},
  year={2026},
  publisher={Elsevier}
}

@ARTICLE{Bou2023ML,
  author={Boualouache, Abdelwahab and Engel, Thomas},
  journal=IEEE_O_CSTO, 
  title={A Survey on Machine Learning-Based Misbehavior Detection Systems for 5G and Beyond Vehicular Networks}, 
  year={2023},
  volume={25},
  number={2},
  pages={1128-1172}}

@ARTICLE{Abdel2025V2X,
  author={Abdel Hakeem, Shimaa A. and Kim, HyungWon},
  journal=IEEE_TITS, 
  title={Advancing Intrusion Detection in V2X Networks: A Comprehensive Survey on Machine Learning, Federated Learning, and Edge AI for V2X Security}, 
  year={2025},
  volume={26},
  number={8},
  pages={11137-11205}}

@ARTICLE{Zhang2023OSN,
  author={Zhang, Xiaoying and Xie, Hong and Yi, Pei and Lui, John C.S.},
  journal=IEEE_TDSC, 
  title={Enhancing Sybil Detection via Social-Activity Networks: A Random Walk Approach}, 
  year={2023},
  volume={20},
  number={2},
  pages={1213-1227}}

@ARTICLE{Al2017OSN,
  author={Al-Qurishi, Muhammad and Al-Rakhami, Mabrook and Alamri, Atif and Alrubaian, Majed and Rahman, Sk Md Mizanur and Hossain, M. Shamim},
  journal={IEEE Access}, 
  title={Sybil Defense Techniques in Online Social Networks: A Survey}, 
  year={2017},
  volume={5},
  number={},
  pages={1200-1219}}

@ARTICLE{Sybil2025Blockchain,
  author={Agrawal, Ankit and Bhatia, Ashutosh and Tiwari, Kamlesh},
  journal={IEEE Open J. Comput. Soc},
  title={Sybil-Resilient Publisher Selection Mechanism in Blockchain-Based MCS Systems}, 
  year={2025},
  volume={6},
  number={},
  pages={586-598}}

@article{patel2025survey,
  title={A Survey of Recent Advancements in Secure Peer-to-Peer Networks},
  author={Patel, Raj and Biswas, Umesh and Kodipaka, Surya and Carroll, Will and Peranich, Preston and Young, Maxwell},
  journal={arXiv preprint arXiv:2509.19539},
  year={2025}
}

@ARTICLE{Ya2024fedSybil,
  author={Yazdinejad, Abbas and Dehghantanha, Ali and Karimipour, Hadis and Srivastava, Gautam and Parizi, Reza M.},
  journal=IEEE_TIFS, 
  title={A Robust Privacy-Preserving Federated Learning Model Against Model Poisoning Attacks}, 
  year={2024},
  volume={19},
  number={},
  pages={6693-6708}}

@ARTICLE{Hammi2022Sybil,
  author={Hammi, Badis and Idir, Yacine Mohamed and Zeadally, Sherali and Khatoun, Rida and Nebhen, Jamel},
  journal=IEEE_TITS, 
  title={Is it Really Easy to Detect Sybil Attacks in C-ITS Environments: A Position Paper}, 
  year={2022},
  volume={23},
  number={10},
  pages={18273-18287}}

@article{benarous2025Pse,
  title={A review of pseudonym change strategies for location privacy preservation schemes in vehicular networks},
  author={Benarous, Leila and Zeadally, Sherali and Boudjit, Saadi and Mellouk, Abdelhamid},
  journal={ACM Comput. Surv.},
  volume={57},
  number={8},
  pages={1--37},
  year={2025},
  publisher={ACM New York, NY}
}

@article{baza2020Sybil,
  title={Detecting sybil attacks using proofs of work and location in vanets},
  author={Baza, Mohamed and Nabil, Mahmoud and Mahmoud, Mohamed MEA and Bewermeier, Niclas and Fidan, Kemal and Alasmary, Waleed and Abdallah, Mohamed},
  journal=IEEE_TDSC,
  volume={19},
  number={1},
  pages={39--53},
  year={2020},
  publisher={IEEE}
}

@article{benadla2022RSSI,
  title={Detecting Sybil attacks in vehicular fog networks using RSSI and Blockchain},
  author={Benadla, Sarra and Merad-Boudia, Omar Rafik and Senouci, Sidi Mohammed and Lehsaini, Mohamed},
  journal=IEEE_TNSM,
  volume={19},
  number={4},
  pages={3919--3935},
  year={2022},
  publisher={IEEE}
}

@article{yao2018RSSI,
  title={Multi-channel based Sybil attack detection in vehicular ad hoc networks using RSSI},
  author={Yao, Yuan and Xiao, Bin and Wu, Gaofei and Liu, Xue and Yu, Zhiwen and Zhang, Kailong and Zhou, Xingshe},
  journal=IEEE_TMS,
  volume={18},
  number={2},
  pages={362--375},
  year={2018},
  publisher={IEEE}
}

@techreport{SAEJ2735_2024,
  author      = {{SAE International}},
  title       = {V2X Communications Message Set Dictionary},
  institution = {SAE International},
  year        = {2024},
  type        = {Standard},
  number      = {SAE J2735},
  address     = {Warrendale, PA, USA},
}

@ARTICLE{Liu2026DTIDS,
  author={Liu, Chang and Zhang, Yurong and Xue, Zheng and Sheng, Zhengguo and Kang, Jiawen and Han, Guojun},
  journal=IEEE_TITS, 
  title={CATwin-IDS: Context-Aware Intrusion Detection System for Both In-Vehicle and External-Vehicle Networks via Digital Twin}, 
  year={2026},
  volume={},
  number={},
  pages={1-15}}

@article{wang2026class,
  title={A class-imbalance-aware intrusion detection system based on spatiotemporal graph neural networks for software-defined vehicles},
  author={Wang, Sishan and Zhao, Youqun and Fu, Xin and Si, Huachao and Wang, Wentao and Xue, Lei},
  journal={J. Inf. Secur. Appl.},
  volume={97},
  pages={104340},
  year={2026},
  publisher={Elsevier}
}

@article{xu2023secure,
  title={Secure intrusion detection by differentially private federated learning for inter-vehicle networks},
  author={Xu, Qian and Zhang, Lei and Ou, Dongxiu and Yu, Wenjuan},
journal={Transp. Res. Rec.},
  volume={2677},
  number={9},
  pages={421--437},
  year={2023}
}

@article{chen2024fast,
  title={Fast and practical intrusion detection system based on federated learning for VANET},
  author={Chen, Xiuzhen and Qiu, Weicheng and Chen, Lixing and Ma, Yinghua and Ma, Jin},
  journal={Comput. Secur.},
  volume={142},
  pages={103881},
  year={2024},
  publisher={Elsevier}
}

@article{aishwarya2026LLM,
  title={Lightweight Misbehavior Detection in the Internet of Vehicles Using Knowledge Distillation from Large Language Models},
  author={Aishwarya, R and Vetriselvi, V and Prahmodh, R and others},
  journal={Comput. Netw.},
  pages={112083},
  year={2026},
  publisher={Elsevier}
}

@ARTICLE{Bin2026GNNSurvey,
  author={Binshaflout, Elham and Hamrouni, Aymen and Ghazzai, Hakim},
  journal=IEEE_TITS, 
  title={Graph Neural Networks for Vehicular Social Networks: Trends, Challenges, and Opportunities}, 
  year={2026},
  volume={27},
  number={2},
  pages={1731-1755}}

@ARTICLE{Feng2026DyGNN,
  author={Feng, ZhengZhao and Wang, Rui and Wang, TianXing and Song, Mingli and Wu, Sai and He, Shuibing},
  journal=IEEE_TKDE, 
  title={A Comprehensive Survey of Dynamic Graph Neural Networks: Models, Frameworks, Benchmarks, Experiments and Challenges}, 
  year={2026},
  volume={38},
  number={1},
  pages={26-46}}

@article{song2022graph,
  title={Graph-based semi-supervised learning: A comprehensive review},
  author={Song, Zixing and Yang, Xiangli and Xu, Zenglin and King, Irwin},
  journal=IEEE_TNNLS,
  volume={34},
  number={11},
  pages={8174--8194},
  year={2022},
  publisher={IEEE}
}

@article{mvula2024SSL,
  title={A Survey on the Applications of Semi-supervised Learning to Cyber-security},
  author={Mvula, Paul Kiyambu and Branco, Paula and Jourdan, Guy-Vincent and Viktor, Herna Lydia},
  journal={ACM Comput. Surv.},
  volume={56},
  number={10},
  pages={1--41},
  year={2024},
  publisher={ACM New York, NY}
}

@inproceedings{tang2025deep,
  title={A Deep Learning Approach to Detecting Multiple Types of Sybil Nodes in VANETs},
  author={Tang, Dong and Mahanti, Aniket and Naha, Ranesh and Pathak, Vishwambhar and Gong, Mingwei},
  booktitle={Proceedings of the 18th IEEE/ACM International Conference on Utility and Cloud Computing},
  pages={1--8},
  year={2025}
}

@INPROCEEDINGS{Chen2022Sybil,
  author={Chen, Ye and Lai, Yingxu and Zhang, Zhaoyi and Li, Hanmei and Wang, Yuhang},
  booktitle={2022 IFIP Networking Conference (IFIP Networking)}, 
  title={Malicious attack detection based on traffic-flow information fusion}, 
  year={2022},
  volume={},
  number={},
  pages={1-9}}

@article{chen2025sybil,
  title={Sybil attack detection and traceability scheme based on temporal heterogeneous graph attention networks},
  author={Chen, Ye and Lai, Yingxu and Zeng, Congai},
  journal={J. Netw. Comput. Appl.},
  volume={242},
  pages={104261},
  year={2025},
  publisher={Elsevier}
}

@article{zhu2024sybil,
  title={Sybil attacks detection and traceability mechanism based on beacon packets in connected automobile vehicles},
  author={Zhu, Yaling and Zeng, Jia and Weng, Fangchen and Han, Dan and Yang, Yiyu and Li, Xiaoqi and Zhang, Yuqing},
  journal={Sensors},
  volume={24},
  number={7},
  pages={2153},
  year={2024}
}

@INPROCEEDINGS{Luo2021Sybil,
  author={Luo, Baiting and Liu, Xiangguo and Zhu, Qi},
  booktitle={2021 IEEE Intelligent Vehicles Symposium (IV)}, 
  title={Credibility Enhanced Temporal Graph Convolutional Network Based Sybil Attack Detection On Edge Computing Servers}, 
  year={2021},
  volume={},
  number={},
  pages={524-531}}

@article{sultana2024Sybil,
  title={Detecting Sybil attacks in VANET: exploring feature diversity and deep learning algorithms with insights into sybil node associations},
  author={Sultana, Rukhsar and Grover, Jyoti and Tripathi, Meenakshi and Sachdev, Manhar Singh and Taneja, Sparsh},
  journal={J. Netw. Syst. Manag.},
  volume={32},
  number={3},
  pages={51},
  year={2024},
  publisher={Springer}
}

@ARTICLE{Abdel2025Sybil,
  author={Abdel Rahman, Naji and Illi, Elmehdi and Althunibat, Saud and Qaraqe, Marwa},
  journal=IEEE_IOTJ, 
  title={Mobility Discloses Genuinity: A Robust Machine Learning-Based Sybil Attack Detection Scheme}, 
  year={2025},
  volume={12},
  number={24},
  pages={53939-53953}}

@ARTICLE{Zhang2023SybilRW,
  author={Zhang, Xiaoying and Xie, Hong and Yi, Pei and Lui, John C.S.},
  journal=IEEE_TDSC, 
  title={Enhancing Sybil Detection via Social-Activity Networks: A Random Walk Approach}, 
  year={2023},
  volume={20},
  number={2},
  pages={1213-1227}}

@ARTICLE{Gong2014,
  author={Gong, Neil Zhenqiang and Frank, Mario and Mittal, Prateek},
  journal=IEEE_TIFS, 
  title={SybilBelief: A Semi-Supervised Learning Approach for Structure-Based Sybil Detection}, 
  year={2014},
  volume={9},
  number={6},
  pages={976-987}}

@ARTICLE{Yang2023SemiSurvey,
  author={Yang, Xiangli and Song, Zixing and King, Irwin and Xu, Zenglin},
  journal=IEEE_TKDE, 
  title={A Survey on Deep Semi-Supervised Learning}, 
  year={2023},
  volume={35},
  number={9},
  pages={8934-8954}}

@article{kristianto2023misbehavior,
  title={Misbehavior detection system with semi-supervised federated learning},
  author={Kristianto, Edy and Lin, Po-Ching and Hwang, Ren-Hung},
  journal={Veh. Commun.},
  volume={41},
  pages={100597},
  year={2023},
  publisher={Elsevier}
}

@article{duan2022app,
  title={Application of a dynamic line graph neural network for intrusion detection with semisupervised learning},
  author={Duan, Guanghan and Lv, Hongwu and Wang, Huiqiang and Feng, Guangsheng},
  journal=IEEE_TITS,
  volume={18},
  pages={699--714},
  year={2022},
  publisher={IEEE}
}

@inproceedings{tian2023sad,
  title={SAD: Semi-supervised anomaly detection on dynamic graphs},
  author={Tian, Sheng and Dong, Jihai and Li, Jinhui and Zhao, Wen and Xu, Xiao and Song, Baokun and Meng, Chang and Zhang, Tianyu and Chen, Lihui},
  booktitle={Proceedings of the 32nd International Joint Conference on Artificial Intelligence (IJCAI)},
  pages={2296--2304},
  year={2023}
}

@article{ekle2024anomaly,
  title={Anomaly detection in dynamic graphs: A comprehensive survey},
  author={Ekle, Ocheme Anthony and Eberle, William},
  journal={ACM Trans. Knowl. Discov. Data},
  volume={18},
  number={8},
  pages={1--44},
  year={2024},
  publisher={ACM New York, NY}
}

@manual{river_doc_overview_2026,
  title = {River API Overview},
  organization = {online-ml/river},
  year = {2026},
  url = {https://riverml.xyz/latest/api/overview/},
  note = {Online; accessed 2026-04-09},
  copyright = {Copyright © 2019-2026}
}

@article{sultana2024coop,
  title={Cooperative approach for data-centric and node-centric misbehavior detection in VANET},
  author={Sultana, Rukhsar and Grover, Jyoti and Tripathi, Meenakshi},
  journal={Vehicular Communications},
  volume={50},
  pages={100855},
  year={2024},
  publisher={Elsevier}
}

@ARTICLE{kamel2020MDS,
  author={Kamel, Joseph and Ansari, Mohammad Raashid and Petit, Jonathan and Kaiser, Arnaud and Jemaa, Ines Ben and Urien, Pascal},
  journal=IEEE_TVT, 
  title={Simulation Framework for Misbehavior Detection in Vehicular Networks}, 
  year={2020},
  volume={69},
  number={6},
  pages={6631-6643},
  doi={10.1109/TVT.2020.2984878}}

@phdthesis{shahid2025securing,
  title={Securing Inter-Vehicular Communications in Connected Vehicles},
  author={Shahid, Muhammad Anwar},
  year={2025},
  school={University of Windsor (Canada)}
}

@ARTICLE{Devika2024VADGAN,
  author={S, Devika and Shrivastava, Rishi Rakesh and Narang, Pratik and Alladi, Tejasvi and Yu, F. Richard},
  journal=IEEE_TVT, 
  title={VADGAN: An Unsupervised GAN Framework for Enhanced Anomaly Detection in Connected and Autonomous Vehicles}, 
  year={2024},
  volume={73},
  number={9},
  pages={12458-12467}}

@inproceedings{cao2006denstream,
  title={Density-based clustering over an evolving data stream with noise},
  author={Cao, Feng and Estert, Martin and Qian, Weining and Zhou, Aoying},
  booktitle={Proceedings of the 2006 SIAM international conference on data mining},
  pages={328--339},
  year={2006},
  organization={SIAM}
}

@article{hahsler2016clustering,
  title={Clustering data streams based on shared density between micro-clusters},
  author={Hahsler, Michael and Bola{\~n}os, Matthew},
  journal=IEEE_TKDE,
  volume={28},
  number={6},
  pages={1449--1461},
  year={2016},
  publisher={IEEE}
}

@STRING{IEEE_TITS        = "{IEEE} Trans. Intell. Transp. Syst."}

@STRING{IEEE_TVT         = "{IEEE} Trans. Veh. Technol."}

@STRING{IEEE_TNNLS         = "{IEEE} Trans. Neural Netw. Learn. Syst."}

@STRING{IEEE_TIFS         = "{IEEE} Trans. Inf. Forensics Security"}

@STRING{IEEE_TKDE        = "{IEEE} Trans. Knowl. Data Eng."}

@STRING{IEEE_TDSC         = "{IEEE} Trans. Dependable Secure Comput."}

@STRING{IEEE_IOTJ        = "{IEEE} Internet Things J."}

@STRING{IEEE_O_CSTO        = "{IEEE} Commun. Surveys Tuts."}

@STRING{IEEE_TNSM        = "{IEEE} Trans. Netw. Ser. Man."}

\begin{IEEEbiographynophoto}{Qian Xu}
is currently a Postdoctoral Fellow with the State Key Laboratory of Internet of Things for Smart City, University of Macau. 
She holds the Ph.D. degree in Transportation Engineering from Tongji University in 2025, the Master’s degree in Traffic Information Engineering and control in 2021, and the bachelor’s degree in Railway Traffic Signal and Control from Lanzhou Jiaotong University  Control from Lanzhou Jiaotong University in 2018. Her research focuses on safety, security and artificial intelligence techniques for autonomous driving and V2X.
\end{IEEEbiographynophoto}

\begin{IEEEbiographynophoto}{Jiaxun Zhang}
is currently pursuing the Ph.D. degree with the State Key Laboratory of Internet of Things for Smart City and the Department of Civil and Environmental Engineering, Faculty of Engineering, University of Macau. She received the M.S. degree in Integrated Sustainable Design from the National University of Singapore in 2023 and the B.E. degree in Traffic Engineering from South China University of Technology in 2022. Her research primarily focuses on the integration of artificial intelligence with autonomous driving technologies and intelligent transportation systems.
\end{IEEEbiographynophoto}

\begin{IEEEbiographynophoto}{Chengyue Wang}
received the BE degree in transportation engineering from Chang’an University in 2021 and the MS degree in civil engineering from the University of Illinois Urbana-Champaign in 2022. He is currently working toward the PhD degree with the State Key Laboratory of Internet of Things for Smart City and the Department of Civil and Environmental Engineering, Faculty of Engineering, University of Macau. His research primarily focuses on the innovative integration of artificial intelligence with autonomous driving technologies and intelligent transportation systems. 
\end{IEEEbiographynophoto}

\begin{IEEEbiographynophoto}{Zhenning Li}(Member, IEEE) 
received his Ph.D. in Civil Engineering from the University of Hawaii at Manoa, Honolulu, Hawaii, USA, in 2019. Currently, he holds the position of Assistant Professor at the State Key Laboratory of Internet of Things for Smart City, as well as the Department of Civil and Environmental Engineering, Faculty of Engineering, and the Department of Artificial Intelligence, Faculty of Information Science and Computing at the University of Macau. Over his academic career, he has published over 80 papers. His main areas of research focus on the intersection of connected autonomous vehicles and Big Data applications in urban transportation systems. He has been honored with several awards, including the TRB best young researcher award and the CICTP best paper award.
\end{IEEEbiographynophoto}

\vfill

\end{document}